\documentclass[11pt, a4paper]{article}
\usepackage{styleBuding}
\usepackage{bm}
\usepackage{color}
\usepackage[usenames,dvipsnames,svgnames,table]{xcolor}
\usepackage{amsmath}
\usepackage{amssymb}
\usepackage{graphicx}
\usepackage{slashed}
\usepackage{soul}
\usepackage{multirow}
\usepackage{subfigure}
\usepackage{indentfirst}
\usepackage{siunitx} % for physics units
\usepackage[utf8]{inputenc}

\pdfoutput=1

\begin{document}
%\maketitle
%\newpage

\title{\boldmath 95 GeV Diphoton and $b \bar{b}$ Excesses
in the General Next-to-Minimal Supersymmetric Standard Model}

%% %simple case: 2 authors, same institution
%% \author{A. Uthor}
%% \author{and A. Nother Author}
%% \affiliation{Institution,\\Address, Country}

% more complex case: 4 authors, 3 institutions, 2 footnotes

%%% authors
\author{Junjie Cao$^{a,b}$, Xinglong Jia$^a$, Jingwei Lian$^c$, Lei Meng$^a$}

\affiliation{ $^a$ Department of Physics, Henan Normal University, Xinxiang 453007, China}
\affiliation{ $^b$  Schools of Physics, Shandong University, Jinan, Shandong 250100, China}
\affiliation{$^c$ Henan Institute of Science and Technology, Xinxiang 453003, China}
%%%
% e-mail addresses: one for each author, in the same order as the authors
\emailAdd{junjiec@alumni.itp.ac.cn}
\emailAdd{JiaXinglong1996@outlook.com}
\emailAdd{lianjw@hist.edu.cn}
\emailAdd{mel18@foxmail.com}

\abstract{The CMS and ATLAS collaborations recently published their results searching for light Higgs bosons, using the complete Run 2 data of the LHC. Both reported an excess in the diphoton invariant mass distribution at $m_{\gamma \gamma} \simeq 95.4~{\rm GeV}$ with compatible signal strengths. The combined result corresponded to a local significance of $3.1\sigma$. Besides, the mass of the diphoton signal coincided with that of the $b\bar{b}$ excess observed at the LEP. Given the remarkable theoretical advantages of the general Next-to-Minimal Supersymmetric Standard Model, we interpret these excesses by the resonant productions of the singlet-dominated CP-even Higgs boson predicted by the theory. Using both analytic formulae and numerical results, we show that the idea can interpret the excesses by broad parameter space without contradicting current experimental restrictions, including those from the 125~{\rm GeV} Higgs data, the dark matter relic abundance and direct detection experiments, and the collider searches for supersymmetry and extra Higgs bosons. Although the explanations are scarcely affected by present Higgs data and the LHC search for supersymmetry, the dark matter physics may leave footprints on them. We also survey the other signals of the light Higgs boson at the LHC. }

\maketitle

\section{Introduction\label{sec:intro}}

The discovery of the Higgs boson at the Large Hadron Collider (LHC) in 2012 proved the existence of a scalar field. It provided essential insights into
the electroweak symmetry breaking (EWSB) and mass generation mechanisms. Although the properties of this boson are consistent with the predictions of the
Standard Model (SM), the notorious hierarchy problem implies that the SM might not be the whole story, and there ought to be a more underlying framework
from physics beyond the SM (BSM) to account for the EWSB. Since the BSM physics often gives rise to extended Higgs sectors in which additional scalar particles
are present, searching for extra Higgs bosons becomes one of the primary objectives of the LHC. In this regard, it is noticeable that the presence of additional
scalars with masses below 125GeV is not excluded if their couplings are suppressed compared to those of the SM Higgs boson. These extra Higgs bosons are within
the reach of the LHC, and with moderately large couplings, they would have been produced in small numbers in past runs. Thus, an intriguing question is whether
there could be hints of one or more additional Higgs bosons in the currently existing searches in the form of no-significant excesses over the background
expectation.

Among the ongoing searches for low-mass Higgs bosons, the CMS collaboration first revealed in 2015 an excess with a local significance of 2.0$\sigma$
in the diphoton invariant-mass distribution at $m_{\gamma \gamma} \simeq 97$ GeV, based on 19.7 fb$^{-1}$ of LHC data at a center-of-mass energy  of
8 TeV~\cite{CMS:2015ocq}. This excess was reinforced to $2.8\sigma$ in 2018 at $m_{\gamma \gamma} \simeq 95$ GeV  after combining the Run 1 data and
35.9 fb$^{-1}$ of LHC data at 13 TeV~\cite{CMS:2018cyk}. Remarkably, CMS released its latest analysis in march 2023, confirming the excess at
$m_{\gamma\gamma}=95.4$ GeV and with a local significance of 2.9$\sigma$ by employing advanced analysis techniques and
utilizing data collected during the first, second, and third years of Run 2, which correspond to integrated luminosities of 36.3 fb$^{-1}$,
41.5 fb$^{-1}$, and 54.4 fb$^{-1}$, respectively, with a shared collision energy of 13 TeV \cite{CMS:2023yay}. By contrast, the observations of
the ATLAS collaboration are somewhat different. Specifically, this collaboration did not find a significant excess around 95~GeV after scrutinizing
80 fb$^{-1}$ of LHC data in 2018~\cite{ATLAS:2018xad}. However, given that its sensitivity was minor, its limits on the diphoton production rate
were not in tension with the CMS results. Encouragingly, the collaboration recently released its analysis in searching for the diphoton resonances
in the mass range from 66~GeV to 110~GeV using the full Run 2 LHC dataset (140~fb$^{-1}$)~\cite{Arcangeletti}. Compared with its previous
one~\cite{ATLAS:2018xad}, using multivariate analysis techniques in background mitigation and event classification improved the sensitivity to
BSM physics. Numerically speaking, this further analysis revealed an excess in the diphoton channel at an invariant mass around 95 GeV and with
a local significance of $1.7\sigma$, remarkably aligning with the  reported CMS findings.

Regarding interpreting the new result from ATLAS and the previously reported ones from CMS, one should note that the observed diphoton events at about 95 GeV might originate from fluctuating a much more extensive background, thus giving rise to a relatively small number of pseudo signals. Therefore, one cannot necessarily expect that the excesses should occur with the same signal strength.   In this context, the phenomenon that both collaborations reported their most significant excess at the same mass value has to be seen as a certain level of coincidence.   However, since for the same mass value, the renormalized diphoton production rates revealed by the two collaborations, namely $\mu^{\rm CMS}_{\gamma \gamma} = 0.33^{+0.19}_{-0.12}$ and $\mu^{\rm ATLAS}_{\gamma \gamma}= 0.18 \pm 0.10$, agree with each other within their uncertainties, it is intriguing to imagine that they arise from the production of a single new particle.   If it proves true, this will be the first sign of new physics in the Higgs-boson sector~\cite{Biekotter:2023oen}. With this assumption, one could obtain a combined signal strength after neglecting possible correlations. It is given by~\cite{Biekotter:2023oen}
\begin{equation}
\mu^{\rm exp}_{\gamma\gamma} \equiv  \mu^{\rm ATLAS + CMS}_{\gamma \gamma} = \frac{\sigma(p p \to \phi \to \gamma\gamma)}{\sigma_{\rm SM} (p p \to H_{\rm SM} \to \gamma\gamma)} = 0.24^{+ 0.09}_{-0.08},  \label{diphoton-rate}
\end{equation}
where $\phi$ is a postulated non-Standard scalar with $m_\phi = 95.4~{\rm GeV}$, responding for the diphoton excess, and $\sigma_{\rm SM}$ denotes the cross
section for a hypothetical SM Higgs boson, $H_{\rm SM}$, at the same mass. According to the analysis in Ref.~\cite{Biekotter:2023oen}, this fitted value corresponds
to a $3.1\sigma$ local excess. We note that the existence of a light $\phi$ might also be hinted at by the results of the Large Electron Positron (LEP)
in 2006, which showed an excess in the $e^+e^- \rightarrow Z\phi\,(\phi\rightarrow b\bar{b})$ mode at $m_{b\bar{b}}\simeq 98$~GeV with a local significance
of $2.3\sigma$ and $\mu_{b \bar b}^\text{exp} = 0.117 \pm 0.057$~\cite{LEPWorkingGroupforHiggsbosonsearches:2003ing,Azatov:2012bz,Cao:2016uwt}. Considering
the limited mass resolution for the dijets at LEP, the $b\bar{b}$ excess could originate from the same particle responsible for the diphoton excess
summarized above\footnote{Another hints on the existence of a scalar with its mass around $95~{\rm GeV}$ include the di-$\tau$ excess reported by
the CMS collaboration~\cite{CMS:2022goy} and discussed below, and the $WW$ excess discussed in Ref.~\cite{Coloretti:2023wng}.}.
We also note that $\mu^{\rm exp}_{\gamma\gamma}$ in Eq.~(\ref{diphoton-rate}) is much smaller than its previous value
$\mu^{\rm CMS}_{\gamma\gamma} = 0.6 \pm 0.2$, derived by the CMS collaboration from the analysis in 2018~\cite{Heinemeyer:2018wzl}.
This situation allows the new particles contributing to the diphoton rate by loop effects to be relatively heavy, consistent with the results
of the LHC search for new states.

The appearance of the excesses mentioned above triggered studies on the possibility to accommodate them in BSM models, which predict a SM-like Higgs boson with
a mass of around 125 GeV and a lighter non-standard Higgs boson.  These models include the SM extensions with a $SU(2)_L$ triplet scalar
field~\cite{Ashanujjaman:2023etj} or vectorlike fermions~\cite{Aguilar-Saavedra:2020wrj, Kundu:2019nqo, Fox:2017uwr}, the two Higgs-doublet
model~\cite{Belyaev:2023xnv, Azevedo:2023zkg, Benbrik:2022dja, Benbrik:2022azi, Haisch:2017gql} and its extensions with an additional real or complex singlet scalar
~\cite{Biekotter:2019mib,Biekotter:2019kde,Biekotter:2020cjs,Biekotter:2021qbc,Biekotter:2021ovi,Heinemeyer:2021msz,Biekotter:2022jyr,Li:2023hsr,
Biekotter:2023oen,Biekotter:2023jld, Aguilar-Saavedra:2023vpd, Banik:2023ecr,Dutta:2023cig},  Radion model~\cite{Sachdeva:2019hvk},
Georgi-Machacek model~\cite{Vega:2018ddp} and the supersymmetric extensions of the SM ~\cite{Fan:2013gjf,Cao:2016uwt,Biekotter:2021qbc,Heinemeyer:2018wzl,
Heinemeyer:2018jcd,Beskidt:2017dil,Abdelalim:2020xfk,Hollik:2020plc,Cao:2019ofo,Biekotter:2019gtq,Biekotter:2017xmf, Choi:2019yrv, Wang:2018vxp,
Domingo:2018uim,Li:2022etb,Ellwanger:2023zjc}. In addition, there were discussions about possible connections of the observed excesses to extra dimensions~\cite{Richard:2017kot},
$B$-anomalies~\cite{Liu:2018xsw}, dark matter (DM)~\cite{Cline:2019okt} and neutrino mass generation mechanisms~\cite{Escribano:2023hxj,Borah:2023hqw}.  Among these models,
the Next-to-Minimal Supersymmetric Standard Model (NMSSM)~\cite{Ellwanger:2009dp} has attracted significant attentions in recent years since it is the most
economical supersymmetric theory to account for the diphoton and $b\bar{b}$ excesses~\cite{Cao:2016uwt,Heinemeyer:2018wzl,Beskidt:2017dil,Choi:2019yrv}.
It augments the popular Minimal Supersymmetric Standard Model (MSSM)
with one gauge-singlet Higgs field $\hat{S}$. Like the MSSM, it provides an elegant solution to the hierarchy problem and realizes the unification of gauge
interactions at a high energy scale. It is distinct in naturally solving  the $\mu$-problem of the MSSM\footnote{In the MSSM, the direct detection
of DM by the LUX-ZEPLIN (LZ) experiment~\cite{LZ:2022ufs} alone has required the higgsino mass to be significantly higher than the electroweak scale,
namely, $\mu \gtrsim 380~{\rm GeV}$~\cite{He:2023lgi}. Although such a large $\mu$ may be generated  by the well-known Giudice-Masiero mechanism in the
gravity-mediated SUSY breaking scenario~\cite{Giudice:1988yz}, it induces severe fine-tuning problems in the light of the LHC Higgs discovery and the
absence of any discovery of supersymmetry when the MSSM runs down from an infrared high energy scale to the electroweak scale~\cite{Arvanitaki:2013yja,
Evans:2013jna,Baer:2014ica}. The NMSSM dynamically generates the $\mu$-parameter of the MSSM after the scalar component field of $\hat{S}$ develops a
vacuum expectation value (vev) of ${\cal{O}}(1~{\rm TeV})$, firstly proposed by P. Fayet~\cite{Fayet:1974pd,Fayet:1976cr}. In this sense, 
the NMSSM is a self-contained supersymmetric theory at the electroweak scale.} and predicting
more feasible DM candidates~\cite{Ellwanger:2014hia,Cao:2021ljw,Cao:2019qng}. As a result, the model's phenomenology is significantly
enriched~\cite{Cao:2013gba,Baum:2017enm,Ellwanger:2018zxt,Cao:2021tuh,Cao:2022ovk}. In addition, concerning the light CP-even Higgs scenario of the model
suited to explain the excesses, the mass of the SM-like Higgs boson may be sizably lifted by both an additional tree-level contribution and
the singlet-doublet mixing effect~\cite{Ellwanger:2011aa,Badziak:2013bda,Cao:2012fz}, which mitigates the significant radiative corrections
from top/stop loops needed to predict the SM-like Higgs boson mass at $125~{\rm GeV}$.

We once studied the capability of the NMSSM with a $\mathbb{Z}_3$ discrete symmetry ($\mathbb{Z}_3$-NMSSM) to explain the excesses~\cite{Cao:2016uwt},
assuming the singlet-dominated CP-even Higgs boson to be responsible for the diphoton and $b\bar{b}$ excesses. One distinct feature of the $\mathbb{Z}_3$-NMSSM
was that there were six input parameters for each of the Higgs sector and the neutralino sector, and four of them, namely $\lambda$, $\kappa$, $\tan \beta$,
and $\mu$, were shared by these two sectors. Consequently, the theory's Higgs and DM physics were entangled. This correlation limited the maximum reach of
the diphoton signal rate after including the restrictions from DM experiments and the 125 GeV Higgs data collected at the LHC, so the theory hardly
explained the diphoton excess at $1\sigma$ level~\cite{Cao:2016uwt}. Noting that the diphoton rate inferred from the excesses has been reduced
significantly, and simultaneously the sensitivities of DM direct detection experiments have been improved by more than one order compared with
the previous study, we recently renewed the research in Ref.~\cite{Cao:2016uwt}. With the same advanced research strategy as that of this work and
the latest relevant experimental results, we found that without finely tuning the model's parameters, the conclusion remained valid, implying
the necessity to loosen the correlation to explain the excesses. This situation could be changed by improving the theory in two directions. One was to augment
the $\mathbb{Z}_3$-NMSSM with the Type-I or inverse seesaw mechanism to generate neutrino masses and take the gauge-singlet sneutrino
as the DM candidate~\cite{Cao:2019ofo}. In this framework, the DM physics was mainly determined by the neutrino Yukawa couplings instead of
the parameters in the Higgs sector. As a result, a wide range of parameter spaces in the Higgs sector were resurrected to be experimentally
allowed and thus enabled the theory to explain the excesses. The other was, motivated by solving the domain wall and tadpole problems of
the $\mathbb{Z}_3$-NMSSM, to neglect the ad hoc $\mathbb{Z}_3$ symmetry and consider the general form of the NMSSM, abbreviated as GNMSSM
in this work~\cite{Ellwanger:2009dp}. In the GNMSSM, although the number of the shared parameters in the Higgs and neutralino sectors remained
four, the Higgs physics was determined by ten parameters. It thus became more flexible and might account for the excesses. Notably, this characteristic
might, in return, enrich the DM properties since the Higgs bosons usually played a role in DM physics.

The excesses in the general NMSSM were analyzed in Ref.~\cite{Choi:2019yrv} by both compact analytic formulas and numerical results.  It concluded that there were parameter spaces
that could explain the excesses without conflicting with the LHC Higgs data. However, such a study did not specify the singlet self-interactions, which could affect Higgs boson masses, and considered the old results of the diphoton excess. More crucially, it neglected the tight restrictions from the DM physics and the
LHC search for supersymmetric particles (sparticles), especially those from the LZ experiment which has reached an unprecedented
sensitivity to the cross sections of spin-independent (SI) and spin-dependent (SD) DM-nucleon scatterings, at the level of $10^{-48} \rm cm^2$
and $10^{-42} \rm cm^2$, respectively~\cite{LZ:2022ufs}. Given that the NMSSM is one of the most popular supersymmetric theories, such
impacts should be included in a comprehensive study of the excesses. This paper will focus on the feasibility of explaining the combined
excess in the GNMSSM and its interplay with the latest DM experimental results.

This paper is organized as follows. In Sec.~\ref{Section-Model}, we briefly introduce the basic skeleton of the GNMSSM and discuss the signal rates of the $\gamma\gamma$ and $b\bar{b}$ events. In Sec.~\ref{Section-excess1}, we perform a comprehensive scan over the model parameter space and present the numerical results by both figures and tables. In Sec.~\ref{Section-implication}, we discuss the implications of the excesses. Finally, the conclusion and comment are made in Sec.~\ref{conclusion}.

\section{Theoretical preliminaries}  \label{Section-Model}
\subsection{The basics of GNMSSM}
The GNMSSM includes the most general renormalizable couplings in its superpotential, given by~\cite{Ellwanger:2009dp}
\begin{eqnarray}
 W_{\rm GNMSSM} = W_{\rm Yukawa} + \lambda \hat{S}\hat{H_u} \cdot \hat{H_d} + \frac{\kappa}{3}\hat{S}^3 + \mu \hat{H_u} \cdot \hat{H_d} + \frac{1}{2} \mu^\prime \hat{S}^2 + \xi\hat{S},  \label{Superpotential}
  \end{eqnarray}
where $W_{\rm Yukawa}$ contains the quark and lepton Yukawa terms in the MSSM superpotential, $\hat{H}_u=(\hat{H}_u^+,\hat{H}_u^0)^T$ and $\hat{H}_d=(\hat{H}_d^0,\hat{H}_d^-)^T$
are the $SU(2)_L$ doublet Higgs superfields, and $\hat{S}$ is the singlet Higgs superfield. $\lambda$ and $\kappa$ are the dimensionless coefficients parameterizing the interactions among the Higgs fields, the same as in the $\mathbb{Z}_3$-NMSSM. The bilinear mass parameters $\mu$ and $\mu^\prime$ and the singlet tadpole parameter $\xi$ describe the $\mathbb{Z}_3$-symmetry violating effects. They are advantageous to solve the tadpole problem~\cite{Ellwanger:1983mg, Ellwanger:2009dp} and the cosmological domain-wall problem of the $\mathbb{Z}_3$-NMSSM~\cite{Abel:1996cr, Kolda:1998rm, Panagiotakopoulos:1998yw}.  Noting that one of these parameters can be eliminated by shifting the $\hat{S}$ field and redefining the other parameters~\cite{Ross:2011xv}, we set $\xi$ to be zero without losing the generality of this study. In this case, the bilinear parameters could stem from an underlying discrete R symmetry, $Z^R_4$ or $Z^R_8$, after the SUSY breaking and might be naturally at the electroweak scale~\cite{Abel:1996cr,Lee:2010gv,Lee:2011dya,Ross:2011xv,Ross:2012nr}. They can significantly change the Higgs and DM physics of the $\mathbb{Z}_3$-NMSSM, which is the focus of this study.

The soft SUSY-breaking Lagrangian for the Higgs fields in the GNMSSM is given by
\begin{eqnarray}
 -\mathcal{L}_{soft} = &\Bigg[\lambda A_{\lambda} S H_u \cdot H_d + \frac{1}{3} \kappa A_{\kappa} S^3+ m_3^2 H_u\cdot H_d + \frac{1}{2} {m_S^{\prime}}^2 S^2 + h.c.\Bigg]  \nonumber \\
& + m^2_{H_u}|H_u|^2 + m^2_{H_d}|H_d|^2 + m^2_{S}|S|^2 ,
  \end{eqnarray}
where $H_u$, $H_d$, and $S$ denote the scalar components of the Higgs superfields, and $m^2_{H_u}$, $m^2_{H_d}$, and $m^2_{S}$ are their soft-breaking masses. After these parameters are fixed by solving the conditional equations to minimize the scalar potential and expressed in terms of the vacuum expectation values of the Higgs fields, $\left\langle H_u^0 \right\rangle = v_u/\sqrt{2}$, $\left\langle H_d^0 \right\rangle = v_d/\sqrt{2}$, and $\left\langle S \right\rangle = v_s/\sqrt{2}$ with $v = \sqrt{v_u^2+v_d^2}\simeq 246~\mathrm{GeV}$, the Higgs sector is described by the following ten free parameters:  $\tan{\beta} \equiv v_u/v_d$, $v_s$, the Yukawa couplings $\lambda$ and $\kappa$, the soft-breaking trilinear coefficients $A_\lambda$ and $A_\kappa$, the bilinear mass parameters $\mu$ and $\mu^\prime$, and their soft-breaking parameters $m_3^2$ and $m_S^{\prime\ 2}$.

In revealing the characteristics of Higgs physics, it is customary to introduce the field combinations of $H_{\rm SM} \equiv \sin\beta {\rm Re}(H_u^0) + \cos\beta {\rm Re} (H_d^0)$, $H_{\rm NSM} \equiv \cos\beta {\rm Re}(H_u^0) - \sin\beta {\rm Re}(H_d^0)$, and $A_{\rm NSM} \equiv \cos\beta {\rm Im}(H_u^0) - \sin\beta  {\rm Im}(H_d^0)$, where $H_{\rm SM}$ stands for the SM Higgs field, and $H_{\rm NSM}$ and $A_{\rm NSM}$ represent the extra doublet fields~\cite{Cao:2012fz}. The elements of $CP$-even Higgs boson mass matrix $\mathcal{M}_S^2$ in the bases $\left(H_{\rm NSM}, H_{\rm SM}, {\rm Re}[S]\right)$ are written as~\cite{Ellwanger:2009dp,Miller:2003ay}
\begin{eqnarray}
  {\cal M}^2_{S, 11}&=& \frac{ \lambda v_s (\sqrt{2} A_\lambda + \kappa v_s + \sqrt{2} \mu^\prime ) + 2 m_3^2  }{\sin 2 \beta} + \frac{1}{2} (2 m_Z^2- \lambda^2v^2)\sin^22\beta, \nonumber \\
  {\cal M}^2_{S, 12}&=&-\frac{1}{4}(2 m_Z^2-\lambda^2v^2)\sin4\beta, \quad {\cal M}^2_{S, 13} = -\frac{\lambda v}{\sqrt{2}} ( A_\lambda + \sqrt{2} \kappa v_s + \mu^\prime ) \cos 2 \beta, \nonumber \\
  {\cal M}^2_{S, 22}&=&m_Z^2\cos^22\beta+ \frac{1}{2} \lambda^2v^2\sin^22\beta,\nonumber  \\
  {\cal M}^2_{S, 23}&=& \frac{\lambda v}{\sqrt{2}} \left[(\sqrt{2} \lambda v_s + 2 \mu) - (A_\lambda + \sqrt{2} \kappa v_s + \mu^\prime ) \sin2\beta \right], \nonumber \\
  {\cal M}^2_{S, 33}&=& \frac{(A_\lambda + \mu^\prime) \sin 2 \beta}{2 \sqrt{2} v_s} \lambda v^2   + \frac{\kappa v_s}{\sqrt{2}} (A_\kappa +  2 \sqrt{2} \kappa v_s + 3 \mu^\prime ) - \frac{\mu}{\sqrt{2} v_s} \lambda v^2, \quad \label{Mass-CP-even-Higgs}
\end{eqnarray}
and those for CP-odd Higgs fields in the bases $\left( A_{\rm NSM}, {\rm Im}(S)\right)$ take the following forms:
\begin{eqnarray}
{\cal M}^2_{P,11}&=& \frac{ \lambda v_s (\sqrt{2} A_\lambda + \kappa v_s + \sqrt{2} \mu^\prime ) + 2 m_3^2  }{\sin 2 \beta}, \quad {\cal M}^2_{P,12} = \frac{\lambda v}{\sqrt{2}} ( A_\lambda - \sqrt{2} \kappa v_s - \mu^\prime ), \nonumber  \\
{\cal M}^2_{P,22}&=& \frac{(A_\lambda + 2 \sqrt{2} \kappa v_s + \mu^\prime ) \sin 2 \beta }{2 \sqrt{2} v_s} \lambda v^2  - \frac{\kappa v_s}{\sqrt{2}} (3 A_\kappa + \mu^\prime) - \frac{\mu}{\sqrt{2} v_s} \lambda v^2 - 2 m_S^{\prime\ 2}. \label{Mass-CP-odd-Higgs}
\end{eqnarray}
After diagonalizing $\mathcal{M}^2_{S}$ and $\mathcal{M}^2_{P}$ with unitary matrices $V$ and $U$, respectively, three $CP$-even and two $CP$-odd Higgs mass eigenstates, denoted as $h_i=\{h,H,h_{\rm s}\}$ and $a_j = \{A_H, A_s\}$, respectively, are obtained:
  \begin{eqnarray}
    h_i & = & V_{h_i}^{\rm NSM} H_{\rm NSM}+V_{h_i}^{\rm SM} H_{\rm SM}+V_{h_i}^{\rm S} Re[S], \quad \quad a_j =  U_{a_j}^{\rm NSM} A_{\rm NSM}+ U_{a_j}^{\rm S} Im [S],
    \label{Vij}
  \end{eqnarray}
where $h$ means the SM-like Higgs boson discovered at the LHC, $H$ and $A_H$ represent heavy doublet-dominated Higgs bosons, and $h_s$ and $A_s$ are singlet-dominated scalars.
The model also predicts a pair of charged Higgs, $H^\pm = \cos \beta H_u^\pm + \sin \beta H_d^\pm$, and its squared mass is
\begin{eqnarray}
    m^2_{H_{\pm}} &=&  \frac{ \lambda v_s (\sqrt{2} A_\lambda + \kappa v_s + \sqrt{2} \mu^\prime ) + 2 m_3^2  }{\sin 2 \beta} + m^2_W - \frac{1}{2}\lambda^2 v^2. \label{Charged Hisggs Mass}
  \end{eqnarray}

The Higgs sector of the GNMSSM has the following features:
  \begin{itemize}
  \item The experimental data have restricted the $H_{\rm NSM}$ and Re$[S]$ components in $h$ to be less than 10\% \cite{ATLAS:2022vkf,CMS:2022dwd}, i.e., $\sqrt{\left (V_h^{\rm NSM} \right )^2 + \left ( V_h^{\rm S} \right )^2} \lesssim 0.1$ and $|V_h^{\rm SM}| \sim 1$. In the limit of $\tan\beta \gg 1$, $h$ is mainly composed of the field Re$[H_u^0]$ and $H$ has the most significant component from Re$[H_d^0]$.
  \item The $CP$-even doublet scalar $H$ almost degenerates with the $CP$-odd scalar $A_H$ and the charged Higgs bosons $H^\pm$ in mass. The LHC searches for extra Higgs bosons combined with the indirect constraints from $B$-physics prefer these bosons to be massive, e.g., $m_{H} \gtrsim 0.5~{\rm TeV}$ \cite{ATLAS:2020zms,CMS:2022goy}.
  \item Current collider data allow the singlet-dominated scalars to be moderately light and contain sizable doublet components~\cite{Cao:2013gba}, which is the starting point of this study.
  \end{itemize}

The neutralino sector in the GNMSSM consists of the bino field $\tilde{B}$, the wino field $\tilde{W}$, the higgsino fields $\tilde{H}_d^0$ and $\tilde{H}_u^0$, and the singlino field $\tilde{S}$. In the bases $\psi \equiv (\tilde{B},\tilde{W},\tilde{H}_d^0,\tilde{H}_u^0,\tilde{S})$, the symmetric neutralino mass matrix takes the following form~\cite{Ellwanger:2009dp}:
  \begin{equation}
    {\cal M} = \left(
    \begin{array}{ccccc}
    M_1 & 0 & -m_Z \sin \theta_W \cos \beta & m_Z \sin \theta_W \sin \beta & 0 \\
      & M_2 & m_Z \cos \theta_W \cos \beta & - m_Z \cos \theta_W \sin \beta &0 \\
    & & 0 & -\mu_{tot} & - \frac{1}{\sqrt{2}} \lambda v \sin \beta \\
    & & & 0 & -\frac{1}{\sqrt{2}} \lambda v \cos \beta \\
    & & & & m_N
    \end{array}
    \right), \label{eq:MN}
    \end{equation}
where $\theta_W$ is the weak mixing angle, and $M_1$ and $M_2$ are the soft-breaking masses of the bino and wino fields, respectively. The higgsino mass $\mu_{tot}$ and the singlino mass $m_N$ are given by $\mu_{tot} \equiv \lambda v_s/\sqrt{2} + \mu$ and  $m_N \equiv  \sqrt{2} \kappa v_s + \mu^\prime$. Diagonalizing $\cal{M}$ by a rotation matrix $N$ then yields five mass eigenstates:
\begin{eqnarray}
\tilde{\chi}_i^0 = N_{i1} \psi^0_1 +   N_{i2} \psi^0_2 +   N_{i3} \psi^0_3 +   N_{i4} \psi^0_4 +   N_{i5} \psi^0_5,
\end{eqnarray}
where $\tilde{\chi}_i^0\,(i=1,2,3,4,5)$ are labeled in a mass-ascending order, and the matrix element $N_{ij}$ parameterizes the component of the field $\psi^0_j$ in $\tilde{\chi}_i^0$. The lightest neutralino $\tilde{\chi}_1^0$ usually acts as a viable DM candidate, and it may be bino-dominated ($\tilde{B}$-dominated) or singlino-dominated ($\tilde{S}$-dominated) to acquire the measured DM relic density~\cite{Baum:2017enm}. One distinct feature of the GNMSSM is that the singlet-dominated scalars may play crucial roles in DM physics if $\tilde{\chi}_1^0$ is the $\tilde{S}$-dominated~\cite{Baum:2017enm,Cao:2021ljw}. Specifically, due to the singlet-doublet coupling and the self-interaction term in the superpotential, these scalars may mediate the DM annihilation and the DM-nucleon scattering. They may also present themselves as the final state of the annihilation.

Notably, so far the physical implications of $A_\kappa$, $\mu$, $\mu^\prime$, $m_3^2$, and $m_S^{\prime\ 2}$ are vague, which motivated us to replace them with the masses of the heavy doublet Higgs fields, the $CP$-even and -odd singlet Higgs fields, and the higgsino and singlino fields, denoted as $m_A \equiv \sqrt{{\cal M}^2_{P,11}}$, $m_B \equiv \sqrt{{\cal M}^2_{S,33}}$, $m_C \equiv \sqrt{{\cal M}^2_{P,22}}$, $\mu_{tot}$, and $m_N$, respectively. With this new set of inputs,  the following identities give $A_\kappa$, $\mu$, $\mu^\prime$, $m_3^2$, and $m_S^{\prime\ 2}$:
\begin{eqnarray}
\mu &= & \mu_{tot} - \frac{\lambda}{\sqrt{2}} v_s, \quad \mu^\prime = m_N - \sqrt{2} \kappa v_s, \quad  m^2_3 = \frac{m^2_A \sin{2\beta}}{2} - \lambda v_s (\frac{\kappa v_s}{2} + \frac{\mu^\prime}{\sqrt{2}} + \frac{A_\lambda}{\sqrt{2}}), \nonumber \\
\kappa A_\kappa &=& \frac{\sqrt{2} m^2_B}{v_s} + \frac{\lambda \mu v^2}{v_s^2}
			- \frac{\lambda(A_\lambda +\mu^\prime) v^2 \sin{2\beta}}{2 v_s^2} - 2 \sqrt{2} \kappa^2 v_s - 3 \kappa \mu^\prime, \nonumber\\
m_S^{\prime 2} &=& -\frac{1}{2} \left[ m^2_C + \frac{\lambda \mu v^2}{\sqrt{2} v_s} + \frac{\kappa v_s (3A_\kappa + \mu^\prime)}{\sqrt{2}} - \frac{\lambda (A_\lambda + 2 \sqrt{2} \kappa v_s+ \mu^\prime) v^2 \sin{2\beta}}{2 \sqrt{2} v_s} \right].   \label{Simplify-1}
\end{eqnarray}
Eqs.~(\ref{Mass-CP-even-Higgs}), (\ref{Mass-CP-odd-Higgs}), and (\ref{Charged Hisggs Mass}) are rewritten as
\begin{eqnarray}
 {\cal M}^2_{S, 11}&=& m_A^2 + \frac{1}{2} (2 m_Z^2- \lambda^2v^2)\sin^22\beta, \quad {\cal M}^2_{S, 12}=-\frac{1}{4}(2 m_Z^2-\lambda^2v^2)\sin4\beta, \nonumber \\
  {\cal M}^2_{S, 13} &=& -\frac{\lambda v}{\sqrt{2}} ( A_\lambda + m_N ) \cos 2 \beta, \quad {\cal M}^2_{S, 22} = m_Z^2\cos^22\beta+ \frac{1}{2} \lambda^2v^2\sin^22\beta, \nonumber  \\
  {\cal M}^2_{S, 23}&=& \frac{\lambda v}{\sqrt{2}} \left[ 2 \mu_{tot} - (A_\lambda + m_N ) \sin2\beta \right], \quad {\cal M}^2_{S, 33} = m_B^2, \quad {\cal M}^2_{P,11} = m_A^2,  \nonumber \\
 \quad {\cal M}^2_{P,22} &=& m_C^2, \quad {\cal M}^2_{P,12} = \frac{\lambda v}{\sqrt{2}} ( A_\lambda - m_N ), \quad  m^2_{H_{\pm}} = m_A^2 + m^2_W -\lambda^2 v^2.  \label{New-mass-matrix}
\end{eqnarray}
In the limit of $\lambda \to 0$, the singlet field will decouple from the doublet Higgs fields, and the field masses can be regarded as physical particle masses to a good approximation. Although this situation can not be directly applied to this study, we verified that adopting the new set of parameters as inputs could significantly boost the process of acquiring the solutions to the excesses compared with the old set of inputs.

\subsection{Formula for the \texorpdfstring{$\gamma\gamma$}{} and \texorpdfstring{$b\bar{b}$}{} signals} \label{Section-excess}
This work assumes $h_s$ to be responsible for the excesses. In the narrow width approximation, the diphoton signal strength normalized to its SM prediction is given by
\begin{eqnarray}
	\mu_{\gamma\gamma}|_{m_{h_s} = 95.4~{\rm GeV}} &=&
  \frac{\sigma_{\rm SUSY}(p p \to h_s)}
       {\sigma_{\rm SM}(p p \to h_s )} \times
       \frac{{\rm Br}_{\rm SUSY}(h_s \to \gamma \gamma)}
       {{\rm Br}_{\rm SM}(h_s \to \gamma \gamma)} \nonumber  \\
& \simeq &  \frac{\sigma_{\rm SUSY, ggF}(pp \to h_s)}{\sigma_{\rm SM, ggF}(pp \to h_s)} \times
       \frac{{\rm \Gamma}_{\rm SUSY}(h_s\to \gamma \gamma)}{{\rm \Gamma}_{\rm SM}(h_s\to \gamma \gamma)} \times \frac{\Gamma_{\rm SM}^{\rm tot}}{\Gamma_{\rm SUSY}^{\rm tot}}  \nonumber \\
& \simeq & \frac{{\rm \Gamma}_{\rm SUSY}(h_s\to g g)}{\Gamma_{\rm SM} (h_s\to g g)} \times
  \frac{{\rm \Gamma}_{\rm SUSY}(h_s\to \gamma \gamma)}{{\rm \Gamma}_{\rm SM}(h_s\to \gamma \gamma)}
   \times \frac{1}{{\rm R}_{\rm Width}},   \nonumber \\
& \simeq & |C_{h_s g g}|^2 \times |C_{h_s \gamma \gamma}|^2 \times  \frac{1}{{\rm R}_{\rm Width}},
  \label{muCMS}
\end{eqnarray}
where the mass of $h_s$ is fixed at $95.4~{\rm GeV}$, the production rate $\sigma( p p \to h_s)$, the decay branching ratio ${\rm Br} (h_s \to \gamma \gamma)$, and the width $\Gamma$, all labeled with the subscript `SUSY',  refer to the predictions of the GNMSSM, and those with the subscript `SM' are acquired by assuming $h_s$ to have SM couplings. Since the gluon fusion (ggF) process is the primary contribution to the Higgs production~\cite{CMS:2023yay,Arcangeletti}, we take $\sigma_{\rm SUSY}(pp \to h_s)/\sigma_{\rm SM}(pp \to h_s) \simeq \sigma_{\rm SUSY, ggF}(pp \to h_s)/\sigma_{\rm SM, ggF}(pp \to h_s) \simeq {\rm \Gamma}_{\rm SUSY}(h_s\to g g)/\Gamma_{\rm SM} (h_s\to g g)$.
$C_{h_s g g}$ is the ratio of the $h_s$-gluon-gluon coupling strengths at the renormalization scale $Q \simeq 100~{\rm GeV}$, $C_{h_s g g} \equiv |{\cal{A}}_{\rm SUSY}^{h_s g g} (Q)/{\cal{A}}_{\rm SM}^{h_s g g} (Q)|$, and thus $\Gamma_{\rm SUSY} (h_s \to g g)/\Gamma_{\rm SM} (h_s \to g g) = |C_{h_s g g}|^2$. The normalized coupling strength of $h_s$ to photons, $C_{h_s \gamma \gamma}$, is similarly defined. Besides, the width $\Gamma_{\rm SUSY}^{\rm tot}$  and the ratio ${\rm R}_{\rm Width} \equiv \Gamma_{\rm SUSY}^{\rm tot}/ \Gamma_{\rm SM}^{\rm tot} $ are acquired by
\begin{eqnarray}
\Gamma_{\rm SUSY}^{\rm tot} &=& \Gamma_{\rm SUSY}(h_s \to b\bar{b}) +  \Gamma_{\rm SUSY}(h_s\to \tau \bar{\tau}) +  \Gamma_{\rm SUSY}(h_s\to c\bar{c}) + \Gamma_{\rm SUSY}(h_s\to g g ) + \cdots \nonumber \\
&=& |C_{h_s b \bar{b}}|^2 \times \Gamma_{\rm SM}(h_s \to b\bar{b}) + |C_{h_s \tau \bar{\tau}}|^2 \times \Gamma_{\rm SM}(h_s \to \tau \bar{\tau}) + |C_{h_s c \bar{c}}|^2 \times \Gamma_{\rm SM}(h_s \to c \bar{c}) \nonumber \\
& & +  |C_{h_s g \bar{g}}|^2 \times \Gamma_{\rm SM}(h_s \to g \bar{g}) + \cdots,  \nonumber \\
{\rm R}_{\rm Width} & = & |C_{h_s b \bar{b}}|^2 \times {\rm Br}_{\rm SM}(h_s \to b\bar{b}) + |C_{h_s \tau \bar{\tau}}|^2 \times {\rm Br}_{\rm SM}(h_s \to \tau \bar{\tau}) + |C_{h_s c \bar{c}}|^2 \times {\rm Br}_{\rm SM}(h_s \to c \bar{c}) \nonumber \\
& & +  |C_{h_s g \bar{g}}|^2 \times {\rm Br}_{\rm SM}(h_s \to g \bar{g}) + \cdots,  \nonumber \\
& \simeq & 0.801 \times |C_{h_s b \bar{b}}|^2 + 0.083 \times |C_{h_s \tau \bar{\tau}}|^2 + 0.041 \times |C_{h_s c \bar{c}}|^2 + 0.067 \times |C_{h_s g \bar{g}}|^2, \label{mugamma}
\end{eqnarray}
respectively, where $C_{h_s f \bar{f}}$ ($f=b$, $\tau$, $c$) are the normalized couplings of $h_s$ to the fermion pair $f \bar{f}$ defined by $C_{h_s f \bar{f}} \equiv {\cal{A}}_{\rm SUSY}^{h_s f \bar{f}} (Q)/{\cal{A}}_{\rm SM}^{h_s f \bar{f}} (Q)$, and the branching ratios in the SM were obtained by the LHC Higgs Cross Section Working Group, which included all known higher-order QCD corrections~\cite{LHCHiggsCrossSectionWorkingGroup:2013rie}.

Similarly, the normalized signal strength of the $b\bar{b}$ excess follows from
\begin{eqnarray}
	\mu_{b\bar{b}}|_{m_{h_s} = 95.4~{\rm GeV}} &=&
  \frac{\sigma_{\rm SUSY}(e^+e^-\to Z h_s)}
       {\sigma_{\rm SM}(e^+e^-\to Z h_s)} \times
       \frac{{\rm Br}_{\rm SUSY}(h_s\to b\bar{b})}
       {{\rm Br}_{\rm SM}(h_s \to b\bar{b})} \nonumber \\
  & = &
  \left|C_{h_s V V}\right|^2 \times |C_{h_s b \bar{b}}|^2 \times \frac{1}{{\rm R}_{\rm Width}}.    \label{muLEP}
\end{eqnarray}
Note that the produced Higgs boson must be CP-even to explain the $b\bar{b}$ excess, since a CP-odd one does not couple to $Z$ boson and thus gives no contributions. By contrast, the scalar may be either CP-even or CP-odd to account for the diphoton excess.

In the GNMSSM, ${\cal{A}}_{\rm SUSY}^{h_s g g}(Q)$ is contributed by the loops mediated by quarks and squarks~\cite{King:2012tr}. The code \textsf{SPheno-4.0.5}~\cite{Porod2003SPheno,Porod2011SPheno3} calculates it by the following formula:
\begin{eqnarray}
{\cal{A}}_{\rm SUSY}^{h_s g g}(Q) &=& \sum_q {\cal{A}}_{\rm SUSY}^{h_s g g, q} (Q) + \sum_{\tilde{q}} {\cal{A}}_{\rm SUSY}^{h_s g g, \tilde{q}} (Q) \nonumber \\
 &=& \sum_q C_{h_s g g} \times {\cal{A}}_{\rm SM}^{h_s q \bar{q}} (Q) \times \frac{{\cal{A}}_{\rm SM}^{h_s g g, q}(Q)}{{\cal{A}}_{\rm SM}^{h_s q \bar{q}} (pole)} + \sum_{\tilde{q}} {\cal{A}}_{\rm SUSY}^{h_s g g, \tilde{q}} (Q)
\end{eqnarray}
where ${\cal{A}}_{\rm SUSY}^{h_s g g, q}$ and ${\cal{A}}_{\rm SUSY}^{h_s g g, \tilde{q}}$ represent the quark and squark contributions, respectively, to the $h_s g g$ coupling in the GNMSSM~\cite{Djouadi:2005gj}, and ${\cal{A}}_{\rm SM}^{h_s g g, q}$ denotes the quark loop contribution in the SM with its expression given in Ref.~\cite{Djouadi:2005gi}. Note that the code has incorporated the higher-order QCD corrections to these quantities by the formulae in Ref.~\cite{Spira:1995rr, Staub:2016dxq}. The $h_s$-quark-quark coupling strength in the SM is given by ${\cal{A}}_{\rm SM}^{h_s q \bar{q}} = m_q/v$~\cite{Djouadi:2005gi}, and ${\cal{A}}_{\rm SM}^{h_s q \bar{q}} (Q)$ and ${\cal{A}}_{\rm SM}^{h_s q \bar{q}} (pole)$ are acquired by quark running and pole masses, respectively. ${\cal{A}}_{\rm SUSY}^{h_s \gamma \gamma}(Q)$ is similarly obtained, except that it receives additional contributions from the W boson, charged Higgs boson, and chargino-mediated loops~\cite{King:2012tr}. We add that the supersymmetric contributions to $C_{h_s g g}$ and $C_{h_s \gamma \gamma}$ are not crucial in this study. Specifically, given the massiveness of the squarks, their contributions to the couplings are typically a few thousandths.  The charginos' contribution to $C_{h_s \gamma \gamma}$ only reaches $1\%$ in an optimum case since $\lambda$ is minor~\cite{Choi:2012he}, and the charged Higgs's contribution is at the level of $0.001\%$.

In this study, we acquired the normalized couplings of $h_s$ to fermions, WW, and ZZ by their tree-level expressions\footnote{The potentially large SUSY-QCD and SUSY-electroweak corrections to the bottom quark Yukawa coupling are minor in this study since gluino and squarks are very massive~\cite{Carena:1999py}.}. After neglecting the difference of the running mass and the pole mass and the supersymmetric contributions, we had the following relations~\cite{Ellwanger:2009dp}:
\begin{eqnarray}
C_{h_s t \bar{t}} &=&  V_{h_s}^{\rm SM} + V_{h_s}^{\rm NSM} \cot \beta  \simeq V_{h_s}^{\rm SM}, \quad C_{h_s b \bar{b}} =  V_{h_s}^{\rm SM} - V_{h_s}^{\rm NSM} \tan \beta,  \quad C_{h_s V V} = V_{h_s}^{\rm SM},  \nonumber \\
C_{h_s c \bar{c}} &=& C_{h_s t \bar{t}}, \quad \quad C_{h_s \tau \bar{\tau}} = C_{h_s b \bar{b}}, \quad \quad C_{h_s g g} \simeq C_{h_s t \bar{t}}, \quad \quad C_{h_s \gamma \gamma} \simeq V_{h_s}^{\rm SM}, \label{hs-couplings}
\end{eqnarray}
where the rotation matrix elements $V^i_j$ are defined in Eq.~(\ref{Vij}). We also concluded $V_{h_s}^{\rm SM} \simeq 0.36$ and $(V_{h_s}^{\rm SM} - V_{h_s}^{\rm NSM} \tan \beta) \simeq 0.70 \times V_{h_s}^{\rm SM} \simeq 0.25 $ (or equivalently, $V_{h_s}^{\rm NSM} \tan \beta \simeq 0.11$)  to acquire the central values of $\mu_{\gamma \gamma}$ and $\mu_{b \bar{b}}$ and the preferred branching ratios to be ${\rm Br}_{\rm SUSY} (h_s \to \gamma \gamma) \simeq 1.86 \times {\rm Br}_{\rm SM} (h_s \to \gamma \gamma) \simeq 2.58 \times 10^{-3}$ and ${\rm Br}_{\rm SUSY} (h_s \to b \bar{b}) \simeq 0.90 \times {\rm Br}_{\rm SM} (h_s \to  b \bar{b}) \simeq 72.6\%$. Alternatively, if we obtained $C_{h_s g g}$ and $C_{h_s \gamma \gamma}$ by the exact formulae, as we always did in this study, these couplings might deviate from $C_{h_s t \bar{t}}$ by $4\%$ and $11\%$, respectively. In this case, we found the central values of $\mu_{\gamma \gamma}$ and $\mu_{b \bar{b}}$ corresponded to $V_{h_s}^{\rm SM} \simeq 0.35$, $(V_{h_s}^{\rm SM} - V_{h_s}^{\rm NSM} \tan \beta) \simeq 0.81 \times V_{h_s}^{\rm SM} \simeq 0.28$, or equivalently, $V_{h_s}^{\rm NSM} \tan \beta \simeq 0.07$, ${\rm Br}_{\rm SUSY} (h_s \to \gamma \gamma) \simeq 1.77 \times {\rm Br}_{\rm SM} (h_s \to \gamma \gamma) \simeq 2.5 \times 10^{-3}$, and ${\rm Br}_{\rm SUSY} (h_s \to b \bar{b}) \simeq 0.95 \times {\rm Br}_{\rm SM} (h_s \to  b \bar{b}) \simeq 76.1\%$. These results reveal that explaining the excesses requires an appropriate $C_{h_s t \bar{t}}$ and simultaneously a relatively suppressed $C_{h_s b \bar{b}}$, which are mainly decided by the Higgs mixings $V_i^j$. Particularly, the small deviations of $C_{h_s g g}$ and $C_{h_s \gamma \gamma}$ from $C_{h_s t \bar{t}}$ can significantly reduce $V_{h_s}^{\rm NSM} \tan \beta$ needed to predict the central values of the excesses, but hardly change $V_{h_s}^{\rm SM}$. The approximations also indicate that any reduction of $V_{h_s}^{\rm NSM} \tan \beta$ can enhance $C_{h_s b \bar{b}}$, leading to the increase of $\mu_{b \bar{b}}$ and the decrease of $\mu_{\gamma \gamma}$
if $C_{h_s g g}$, $C_{h_s \gamma \gamma}$, and $C_{h_s V V}$ are fixed (note that $C_{h_s g g}$ and $C_{h_s \gamma \gamma}$ are insensitive to $C_{h_s b \bar{b}}$, and $C_{h_s V V}$ is independent of $C_{h_s b \bar{b}}$). This property explains the prediction of $\mu_{b \bar{b}} > 0.117$ for $\mu_{\gamma \gamma} = 0.24$, frequently encountered in this study and shown in Fig.~\ref{Fig11}.

Furthermore, we point out that explaining the diphoton and $b \bar{b}$ excesses nontrivially restricts the parameters in the Higgs sector. Specifically, after neglecting the renormalization group running of the input parameters and the radiative corrections to ${\cal{M}}_S^2$ in Eq.~(\ref{New-mass-matrix}), one can express the eigenstate equations of $h_s$ as follows:
\begin{eqnarray}
\sum_{j={\rm NSM},{\rm SM},{\rm S}} \left [ {\cal{M}}_S^2 \right ]^i_j V_{h_s}^j = m_{h_s}^2 V_{h_s}^i.
\end{eqnarray}
Noting $m_{h_s}, m_h \ll m_A$, implying that the mixings of $H_{\rm NSM}$ with $H_{\rm SM}$ and $Re[S]$ are small, we acquire the following approximations:
\begin{eqnarray}
V_{h_s}^{\rm NSM} &\simeq & - \frac{V_{h_s}^S}{\sqrt{2}} \times \frac{\lambda (A_\lambda + m_N) v}{m_A^2}, \quad \lambda \mu_{tot} \simeq \frac{V_{h_s}^{\rm SM} V_{h_s}^{\rm S}}{\sqrt{2}} \times \frac{m_{h_s}^2 - m_h^2}{v}, \nonumber \\
m_B^2 & \simeq & m_{h_s}^2 |V_{h_s}^S|^2 + m_h^2 |V_{h_s}^{\rm SM}|^2,  \quad {\cal{M}}_{S,22}^2 \simeq m_h^2 |V_{h_s}^{\rm S}|^2 + m_{h_s}^2 |V_{h_s}^{\rm SM}|^2, \label{Approximation-relations}
\end{eqnarray}
in the large $\tan \beta$ limit. These expressions indicate that the observed excesses have restricted $m_B$ and ${\cal{M}}_{S,22}$ within narrow ranges. They also suggest
\begin{eqnarray}
\lambda \simeq 0.06 \times \left ( \frac{V_{h_s}^{\rm SM}}{0.35} \right ) \times \left ( \frac{\mu_{tot}}{100~{\rm Gev}} \right )^{-1}, \label{Approximation-relations-1}
\end{eqnarray}
and
\begin{eqnarray}
\lambda \gtrsim 0.014 \times \left ( \frac{\tan \beta}{30} \right )^{-1} \times  \left ( \frac{A_\lambda + m_N }{1~{\rm TeV}} \right )^{-1} \times \left ( \frac{m_A}{1~{\rm TeV}} \right )^2,  \label{Approximation-relations-2}
\end{eqnarray}
to predict $V_{h_s}^{\rm NSM} \tan \beta \gtrsim 0.07$. Since that the LHC searches for electroweakinos have set $\mu_{tot} \gtrsim 200~{\rm GeV}$~\cite{ATLAS:2021moa}, one can infer that explaining the excesses prefers $\lambda \lesssim 0.03$, and consequently $\tan \beta/30 \times (A_\lambda + m_N)  \gtrsim 2~{\rm TeV}$ for $m_A = 1~{\rm TeV}$.

\section{Explanation of the excesses \label{Section-excess1}}
This section introduces our sampling strategy and explains the excesses based on numerical results.  We utilized the package \textsf{SARAH-4.14.3}~\cite{SARAH_Staub2008,SARAH3_Staub2012,SARAH4_Staub2013,SARAH_Staub2015} to build the model routines of the GNMSSM and the codes \textsf{SPheno-4.0.5}~\cite{Porod2003SPheno,Porod2011SPheno3} and \textsf{FlavorKit}~\cite{Porod:2014xia} to generate particle spectrum and compute low energy flavor observables, respectively. We calculated the DM physics observables with the package \textsf{MicrOMEGAs-5.0.4}~\cite{Belanger2002,Belanger2004,Belanger2005,Belanger2006,BelangerRD2006qa,Belanger2008,Belanger2010pz,Belanger2013,Barducci2016pcb,Belanger2018}. We analyzed the acquired samples using the posterior probability density function (PDF) in Bayesian inference and the profile likelihood (PL) in Frequentist statistics~\cite{Fowlie:2016hew}.

\subsection{Research strategy}
\begin{table}[tbp]
\caption{The parameter space explored in this study, assuming all the inputs were flatly distributed in prior since they have clear physical meanings. Considering that the soft trilinear coefficients for the third-generation squarks, $A_t$ and $A_b$, could significantly affect the SM-like Higgs boson mass by radiative corrections, we took $A_t = A_b$ and varied them. The unmentioned dimensional SUSY parameters were not crucial to this study, so we fixed $M_3 =3~{\rm TeV}$, $m_C = 800~{\rm GeV}$, and a shared value of $2~{\rm TeV}$ for the others to be consistent with the LHC search for new physics. We defined all these parameters at the renormalization scale $Q_{inp} = 1~{\rm TeV}$ and {\bf acquired the space by several trial scans over much broader parameter spaces}.
\label{tab:1}}
\centering

\vspace{0.3cm}

\resizebox{0.7\textwidth}{!}{
\begin{tabular}{c|c|c|c|c|c}
\hline
Parameter & Prior & Range & Parameter & Prior & Range   \\
\hline
$\lambda$ & Flat & $0 \sim 0.03$ & $\kappa$ & Flat & $-0.2 \sim 0.2$  \\
$\tan \beta$ & Flat & $5 \sim 60$ & $v_s/{\rm TeV}$ & Flat & $0.1 \sim 1.0 $ \\
$A_t/{\rm TeV}$ & Flat & $1.0 \sim 3.0$ & $A_\lambda/{\rm TeV}$ & Flat & $ 0 \sim 2.0$ \\
$m_A/{\rm TeV}$ & Flat & $0.6 \sim 2.0$ &$m_B/{\rm GeV}$ & Flat & $90 \sim 120$ \\
$\mu_{\rm tot}/{\rm TeV}$ & Flat & $0.4 \sim 1.0$ & $m_N/{\rm TeV}$ & Flat & $-1.0 \sim 1.0$   \\
$M_1/{\rm TeV}$ & Flat & $-1.0 \sim -0.2$ & $M_2/{\rm TeV}$ & Flat & $0.3 \sim 1.0$ \\
\hline
\end{tabular}}
\end{table}

We performed a sophisticated scan over the parameter space in Table~\ref{tab:1}, using the MultiNest algorithm with ${\it{nlive}} = 16000$~\cite{MultiNest2009}\footnote{{\it{Nlive}} in the MultiNest method signifies the number of active or live points determining the iso-likelihood contour in each iteration~\cite{MultiNest2009,Importance2019}.
 The larger it is, the more detailed the scan process will be in surveying the parameter space.}.
We constructed the following likelihood function to guide the scan:
\begin{eqnarray}
\mathcal{L} & \equiv & \mathcal{L}_{\gamma \gamma + b \bar{b}} \times \mathcal{L}_{\rm Res} = \exp \left [-\frac{\chi^2_{\gamma \gamma +b \bar{b}}}{2} \right ] \times  \mathcal{L}_{\rm Res} \nonumber \\
&=&  \exp\left\{ -\frac{1}{2}\left [ \left( \frac{\mu_{\gamma\gamma} - 0.24}{0.08}\right)^2 + \left( \frac{\mu_{b\bar{b}} - 0.117}{0.057}\right)^2\right] \right \}_{m_{h_s} \simeq 95{\rm GeV}} \times \mathcal{L}_{\rm Res}, \label{chi2-excesses}
\end{eqnarray}
where $\mathcal{L}_{\rm Res}$ represented the restrictions from pertinent experiments on the theory: $\mathcal{L}_{\rm Res} = 1$ by our definition if the limitations were satisfied, and otherwise, $\mathcal{L}_{\rm Res} = \exp\left [-100 \right ]$. These restrictions included:

\begin{figure*}[t]
		\centering
%		\resizebox{0.8\textwidth}{!}{
%        	\includegraphics{fig4.pdf}
%        	}
\includegraphics[width=1\linewidth]{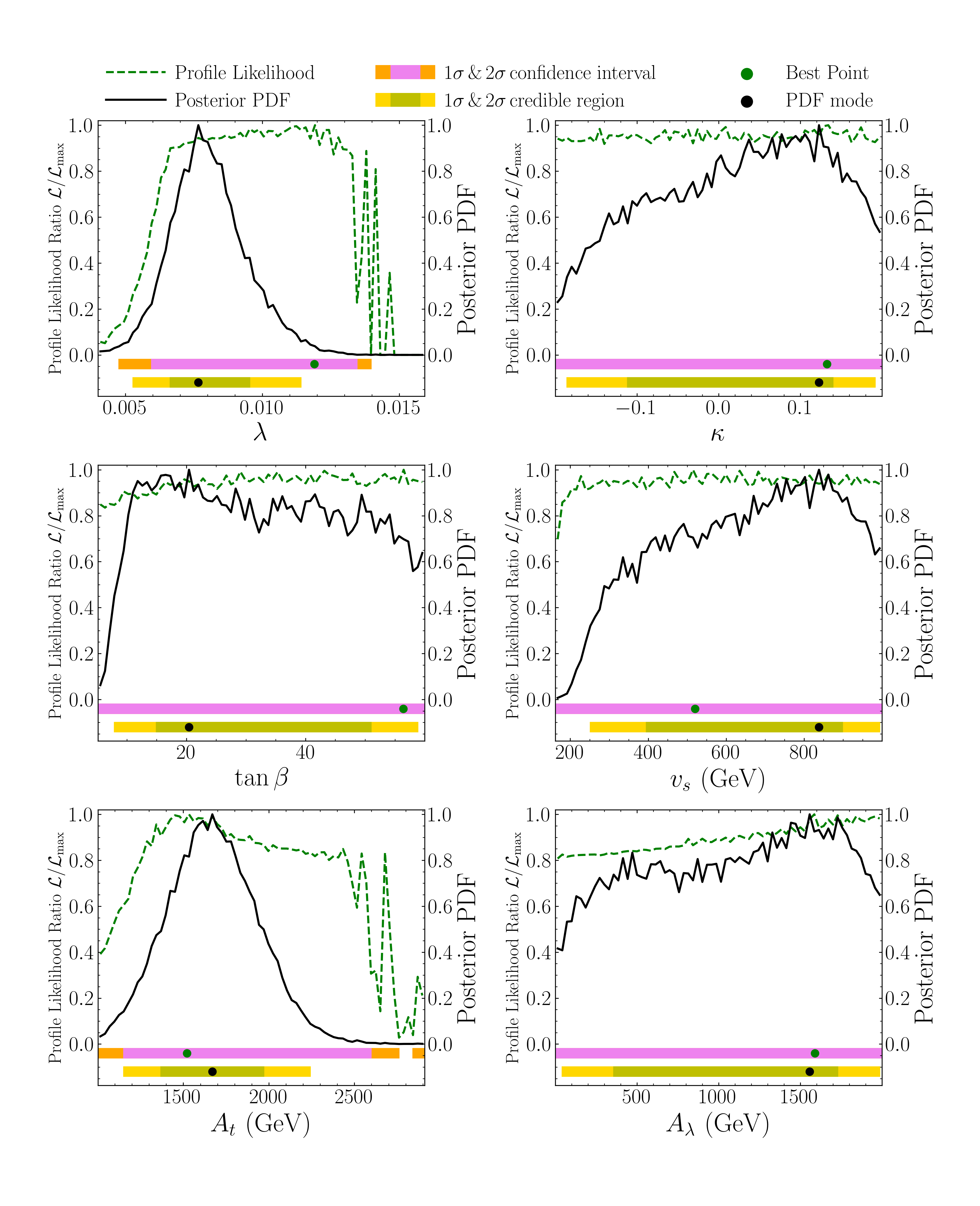}

\vspace{-1.2cm}

\caption{ One-dimensional profile likelihoods (dashed line) and posterior probability density functions (solid line) of the input parameters $\lambda$, $\kappa$, $\tan \beta$, $v_s$, $A_t$, and $A_\lambda$ for the $\tilde{B}$-dominated DM case. The violet and orange bands show the $1\sigma$ and $2\sigma$ confidence intervals, respectively, and the green dot marks the best point corresponding to $\chi^2_{\gamma \gamma + b\bar{b}} =0.27$. The yellow  and golden bands denote the $1\sigma$ and $2\sigma$ credible regions and the black dot denotes the mode of the posterior probability density function. All these statistical measures have been briefly introduced in Ref.~\cite{Fowlie:2016hew}. Since we focus on the regions where both the profile likelihoods and the posterior distributions are large, the plotted ranges of the inputs are usually narrower than those listed Table~\ref{tab:1}.  \label{Fig1}}
\end{figure*}

\begin{figure*}[t]
		\centering
%		\resizebox{0.8\textwidth}{!}{
%        	\includegraphics{fig4.pdf}
%        	}
\includegraphics[width=1\linewidth]{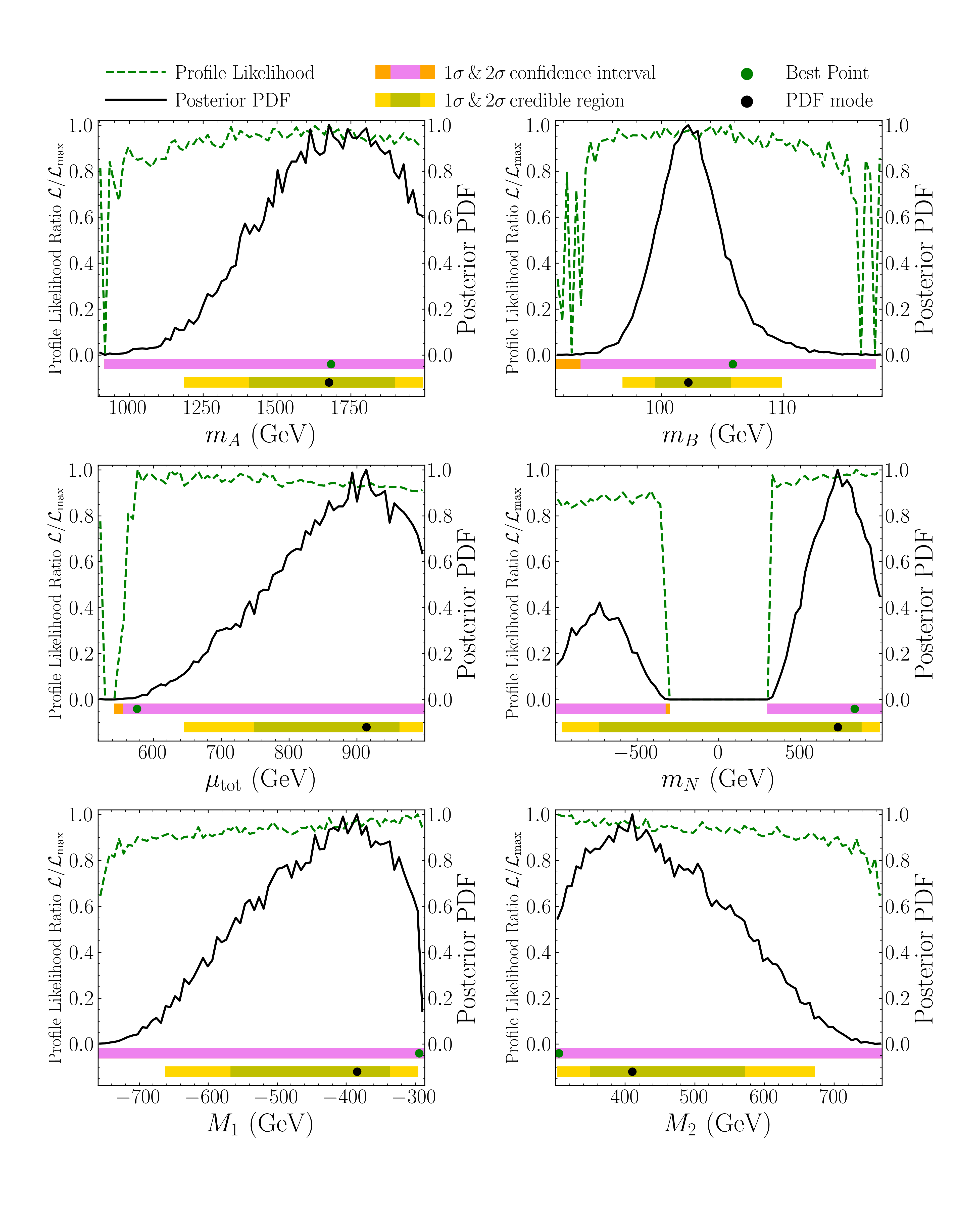}

\vspace{-1.2cm}

\caption{Same as Fig.~\ref{Fig1}, but for the inputs $m_A$, $m_B$, $\mu_{tot}$, $m_N$, $M_1$, and $M_2$.  \label{Fig2} }
\setlength{\abovecaptionskip}{-2.0cm}
\end{figure*}

\vspace{-0.1cm}

\begin{itemize}
\item A proper $h_s$ mass range to explain the excesses: $94.4~{\rm GeV} \leq m_{h_s} \leq 96.4~{\rm GeV}$, where we assumed $1~{\rm GeV}$ theoretical and experimental uncertainties in determining $m_{h_s}$.
\item Higgs data fit. Given that $h$ corresponded to the LHC-discovered Higgs boson, its properties should be consistent with the Higgs measurements by the ATLAS and CMS collaborations at the $95\%$ confidence level. A p-value larger than 0.05 was essential, which was tested by the code \textsf{HiggsSignal-2.6.2}~\cite{HS2013xfa,HSConstraining2013hwa,HS2014ewa,HS2020uwn}.
\item Direct searches for extra Higgs bosons at LEP, Tevatron and LHC. This requirement was implemented by the code \textsf{HiggsBounds-5.10.2}~\cite{HB2008jh,HB2011sb,HBHS2012lvg,HB2013wla,HB2020pkv}.
\item Pertinent DM relic density, $0.096 \leq \Omega h^2 \leq 0.144$. We took the central value of $\Omega {h^2}=0.120$ from the Planck-2018 data~\cite{Planck:2018vyg} and assumed  theoretical uncertainties of $20\%$ in the density calculation.
\item DM direct detection bounds from the LZ experiments on both the SI DM-nucleon scattering cross section, $\sigma_p^{\rm SI}$, and the SD one, $\sigma_n^{\rm SD}$~\cite{LZ:2022ufs}. The DM indirect searches from the observation of dwarf galaxies by the Fermi-LAT collaboration were not included since they had no restrictions on the GNMSSM when $|m_{\tilde{\chi}_1^0}| \gtrsim 100~{\rm GeV}$~\cite{Fermi-LAT:2015att}.
	\item $B$-physics observables. The branching ratios of $B_s \to \mu^+ \mu^-$ and $B \to X_s \gamma$ should be consistent with their experimental measurements at the $2\sigma$ level~\cite{pdg2018}.
    \item Vacuum stability. The vacuum state of the Higgs potential should be either stable or long-lived. This condition was tested by the code \textsf{Vevacious}~\cite{Camargo-Molina:2013qva}.
\end{itemize}

\begin{figure}[t]
	\centering
	\includegraphics[width=1\linewidth]{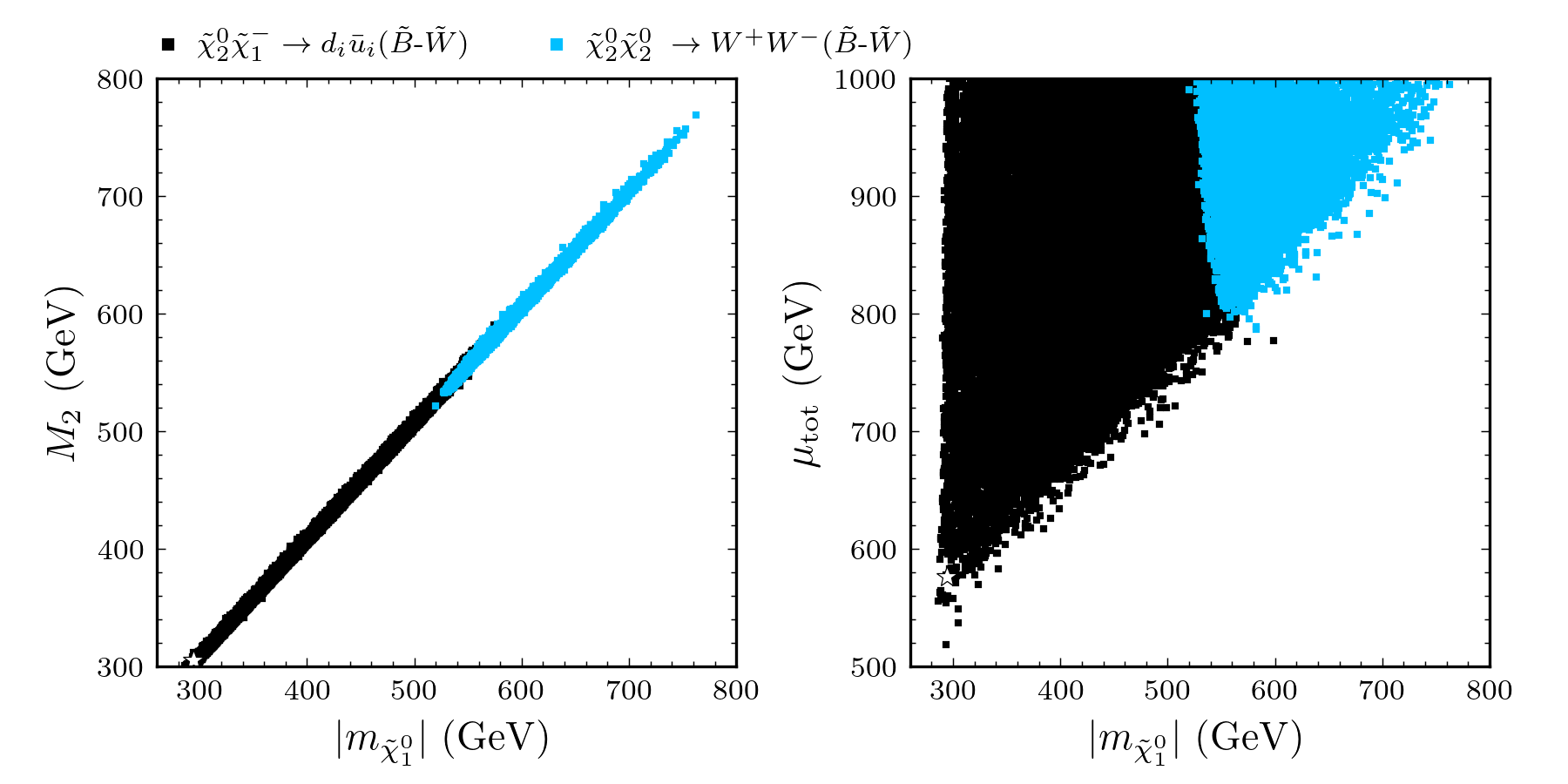}

\vspace{-0.3cm}

\caption{Scattering plots of the samples predicting the $\tilde{B}$-dominated DM, projected onto $M_2-|m_{\tilde{\chi}_1^0}|$ and $\mu_{tot}-|m_{\tilde{\chi}_1^0}|$ planes, respectively.  The left panel reveals that the DM achieved the measured relic abundance by co-annihilating with the wino-like electroweakinos. The largest contribution to the abundance comes from the channels $\tilde{\chi}_2^0 \tilde{\chi}_1^- \to d_i \bar{u}_i$ (i=1,2,3 denote the quark generations) for the black samples and the channel $\tilde{\chi}_2^0 \tilde{\chi}_2^0 \to W^+ W^-$ for the blue samples. The right panel indicates that the LZ results have set a lower bound of $520~{\rm GeV}$ on $\mu_{tot}$ for $|m_{\tilde{\chi}_1^0}|=300~{\rm GeV}$, given that the DM-nucleon scattering cross sections for the $\tilde{B}$-dominated DM are inversely proportional to $\mu_{tot}^2$~\cite{Cao:2019qng}. This bound monotonously increases as the DM becomes heavier. \label{Fig3}}
\end{figure}

In this study, we were particularly interested in the samples that could explain the diphoton and $b\bar{b}$ excesses at the $2 \sigma$ level and be consistent with all the restrictions.  We decided whether they passed the limitations from the LHC search for the electroweakinos by the program \textsf{SModelS-2.1.1}, which encoded various event-selection efficiencies by the topologies of SUSY signals~\cite{Khosa:2020zar}. Moreover, noting that the exclusion capability of this program on the samples was limited by its database and strict working prerequisites, we further surveyed some cases by simulating the analyses listed in Tables 1 of Ref.~\cite{Cao:2022ovk} and the research in Ref.~\cite{ATLAS:2021yqv}\footnote{ ATLAS searched the electroweakino productions by the fully hadronic final states~\cite{ATLAS:2021yqv}, acquiring the hitherto tightest bounds on the electroweakino mass. We encoded this analysis into the package \texttt{CheckMATE-2.0.26}~\cite{Drees:2013wra,Dercks:2016npn, Kim:2015wza} and validated our implementation.}, adopting the strategy of Ref.~\cite{Cao:2022ovk}.  We concluded that the LHC restriction had no impact on the results.  We will explain this phenomenon later.

\vspace{-0.2cm}

\subsection{Numerical Results}

We acquired about 64 thousand samples that were consistent with the experimental restrictions. Analyzing the properties of the samples indicated that the DM might be $\tilde{B}$-dominated or $\tilde{S}$-dominated if one attempted to explain the excesses at the $2\sigma$ level, and they contributed to the total Bayesian evidence by $47\%$ and $53\%$, respectively. In the following, we studied these two types of samples separately. We began with the popular $\tilde{B}$-dominated DM case, but we would show that the $\tilde{S}$-dominated DM case was superior to it in predicting a smaller $\chi^2_{\gamma \gamma + b\bar{b}}$ for the best point.

%\vspace{-0.5cm}

\subsubsection{Bino-dominated DM case}

 We first studied the distributions of various theoretical parameters, including the one-dimensional PL and posterior PDF\footnote{The frequentist PL is the most significant likelihood value in a specific parameter space~\cite{Fowlie:2016hew}. Given a set of input parameters $\Theta \equiv (\Theta_1,\Theta_2,\cdots)$, one can acquire the one-dimensional PL by changing the other parameters to maximize the likelihood function, i.e.,
	\begin{eqnarray}
\mathcal{L}(\Theta_A)=\mathop{\max}_{\Theta_1,\cdots,\Theta_{A-1},\Theta_{A+1},\cdots}\mathcal{L}(\Theta), \nonumber
	\end{eqnarray}
The PL reflects the preference of a theory on the parameter space. For a given point $\Theta_A$, it represents the capability of the point in the theory to account for experimental data. By contrast, the one-dimensional posterior PDF is obtained by integrating the posterior PDF from the Bayesian theorem, $P(\Theta)$, over the rest of the model inputs:
\begin{eqnarray}
	P(\Theta_A)&=&\int{P(\Theta) d\Theta_1 d\Theta_2 \cdots d\Theta_{A-1} d\Theta_{A+1} \cdots  \cdots }.  \nonumber
\end{eqnarray}
It reflects the preference for the samples acquired in the scan. We emphasize that these definitions can be extended to high-dimensional distributions without changing their meanings.
We also emphasize that these statistical measures depend on the studied parameter space, and the posterior PDF also depends on the prior distributions of the input parameters. The intrinsic physics of the considered theory determine all these quantities.} of $\lambda$, $\kappa$, $\tan \beta$, $v_s$, $A_t$, and $A_\lambda$ shown in Fig.~\ref{Fig1}, those of $m_A$, $m_B$, $\mu_{tot}$, $m_N$, $M_1$, and $M_2$ in Fig.~\ref{Fig2}, and the scattering plots of the samples projected onto $M_2-|m_{\tilde{\chi}_1^0}|$ and $\mu_{tot}-|m_{\tilde{\chi}_1^0}|$ planes, respectively, in Fig.~\ref{Fig3}.
The PL distributions indicate that the GNMSSM can explain the diphoton and $b\bar{b}$ excesses at the $2\sigma$ level in broad parameter space except that $m_B$ and $\lambda$ are restricted within narrow ranges. As noted in the approximations in Eqs.~(\ref{Approximation-relations}-\ref{Approximation-relations-1}), the underlying reasons are as follows:
\begin{itemize}
\item Since the signal rates of the excesses have determined $V_{h_s}^{\rm S}$ and $V_{h_s}^{\rm SM}$, $m_B$ is fixed by the relation $m_B^2 \simeq  m_{h_s}^2 |V_{h_s}^S|^2 + m_h^2 |V_{h_s}^{\rm SM}|^2$, leading to $90~{\rm GeV} \lesssim m_B \lesssim 118~{\rm GeV}$;
\item Barring no fine tunings, only the case characterized by a small $\lambda$ and a sufficiently large $\mu_{tot}$ can predict an appropriate $V_{h_s}^{\rm SM}$ to explain the excesses and simultaneously coincide with the results of the DM direct detection experiments and the LHC search for electroweakinos. We will discuss the LHC constraints later.
\end{itemize}
We emphasize that this small $\lambda$ significantly suppresses $V_{h_s}^{\rm NSM}$. Consequently a sufficiently large $(A_\lambda + m_N)$ must be present to enhance $V_{h_s}^{\rm NSM}$, trying to satisfy the condition of $V_{h_s}^{\rm NSM} \tan \beta \simeq 0.10$ needed to acquire the central values of the excesses. This feature explains why the PL of $A_\lambda$ monotonously increases as $A_\lambda$ becomes large. Besides, $m_A$ is restricted from about $1~{\rm TeV}$ to $ 2~{\rm TeV}$, where the lower bound mainly originates from the LHC searches for extra Higgs bosons and B-physics, and the upper bound arises from the boundary of the explored parameter space in Table~\ref{tab:1}.

\begin{figure}[t]
	\centering
	\includegraphics[width=1.0\linewidth]{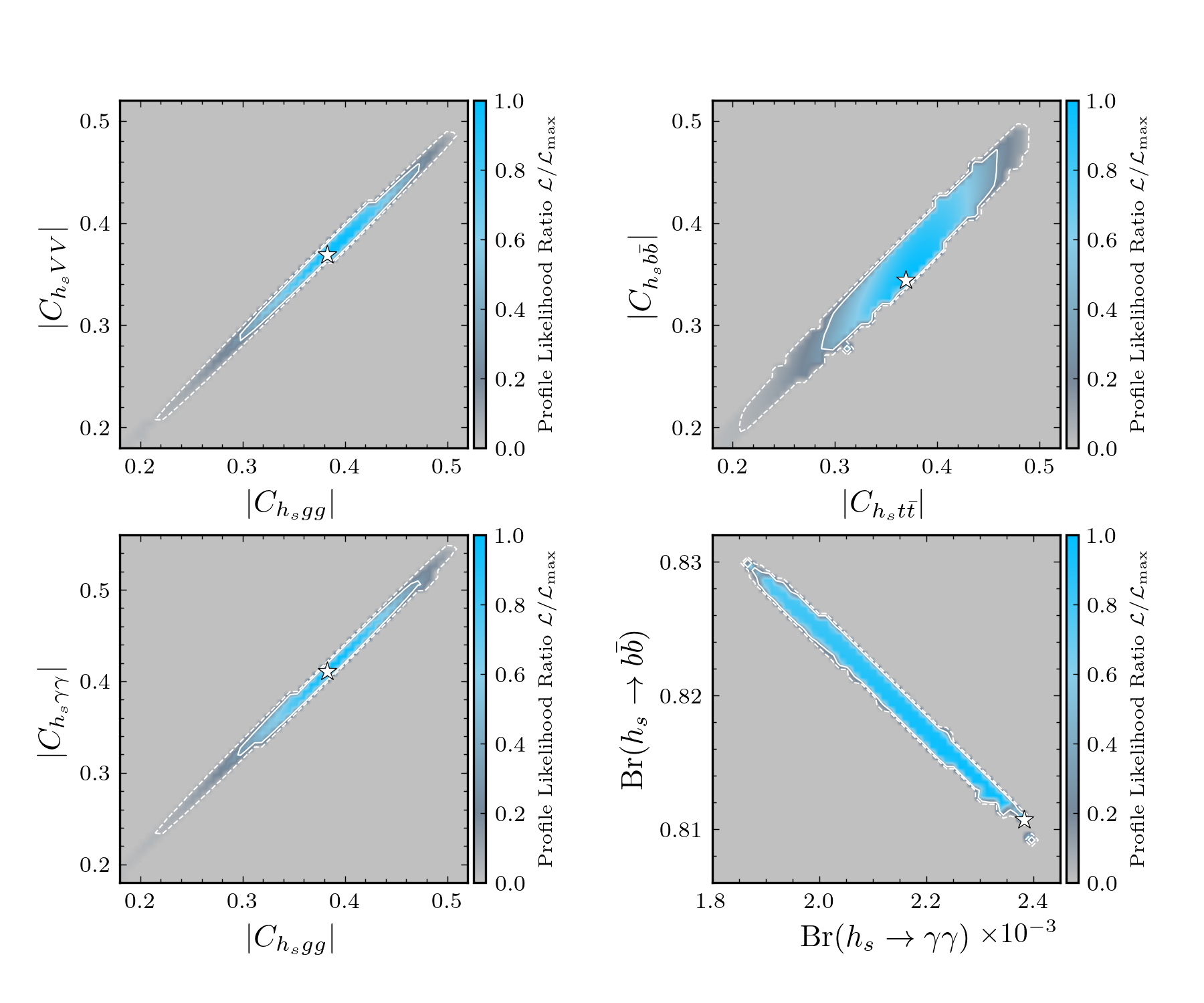}

\vspace{-0.7cm}

\caption{Two-dimensional profile likelihood map of $\mathcal{L}_{\gamma \gamma + b \bar{b}}$ in Eq.~(\ref{chi2-excesses}), projected onto $|C_{h_s V V}|-|C_{h_s g g}|$,  $|C_{h_s b \bar{b}}|-|C_{h_s t \bar{t}}|$, $|C_{h_s \gamma \gamma}|-|C_{h_s g g}|$, and $Br(h_s \to \gamma \gamma) - Br(h_s \to b \bar{b})$ planes, respectively. The best point is marked with a star symbol. Its $\chi^2_{\gamma \gamma + b \bar{b}}$ is equal to 0.27 and its other information are presented in Table~\ref{tab:2} as P1 point. The boundaries for $1\sigma$ and $2\sigma$ confidence intervals correspond to $\chi^2_{\gamma \gamma + b \bar{b}} = 2.3$ and $\chi^2_{\gamma \gamma + b \bar{b}} = 6.18$, respectively, which are labeled as solid and dashed lines.  \label{Fig4}}
\end{figure}

The posterior PDFs and the scattering plots reveal the following physics:
\begin{itemize}
\item The DM physics shown in Fig.~\ref{Fig3} prefers relatively small $|M_1|$ and $M_2$, $|M_N| > |M_1|$, and simultaneously a large $\mu_{tot}$. It also prefers a negative $M_1$ to suppress the SI DM-nucleon scattering by canceling different contributions~\cite{Cao:2019qng}. Furthermore, since the $\tilde{B}$-dominated DM case is featured by a small $\lambda$ and heavy higgsinos, the charginos' contribution to $C_{h_s \gamma \gamma}$ is always less than $1\%$~\cite{Choi:2012he}.
\item Although $V_{h_s}^{\rm NSM}$ depends on $A_\lambda$ and $m_A$, $V_{h_s}^{\rm NSM} \tan \beta \gtrsim 10^{-2}$ preferred to explain the excesses at the $2\sigma$ level can not tightly restrict  $A_\lambda$, $m_A$, and $\tan \beta$ because this condition can be readily achieved by multiple ways, given that the observed excess rates are not extraordinarily large and simultaneously their theoretical predictions are coherently related. One may understand this point by considering the extreme case where $m_A$ in Eq.~(\ref{Approximation-relations}) is tremendously large or $(A_\lambda + m_N)$ vanishes, leading to $V_{h_s}^{\rm NSM} \simeq 0$, and consequently $C_{h_s b \bar{b}} \simeq C_{h_s t \bar{t}} \simeq C_{h_s g g} \simeq C_{h_s \gamma \gamma}$, and  $\mu_{b \bar{b}} \simeq \mu_{\gamma \gamma}$. Then, the theoretical predictions of $\mu_{\gamma \gamma} = \mu_{b \bar{b}} = 0.117$ and $\mu_{b \bar{b}} = \mu_{\gamma \gamma} = 0.24$, easily achieved in the GNMSSM, correspond to $\chi^2_{\gamma \gamma + b \bar{b}} = 2.4$ and $4.7 $, respectively, implying that this case explains well the excesses.

    In addition, it is noticeable that $m_N$ tends to be of the same sign as $A_\lambda$ to avoid a significant cancelation in contributing to $V_{h_s}^{\rm NSM}$. It is beneficial to acquire the central values of the excesses.
\item The trilinear coefficient $A_t$ significantly affects the SM-like Higgs boson mass $m_h$, the mixing $V_{h_s}^{\rm SM}$, and the $h_s g g$ and $h_s \gamma \gamma$ coupling strengths via $\tilde{t}$-mediated loops. Given that we have fixed the soft-breaking masses of the squarks at $2~{\rm TeV}$, $A_t$s around $1.7~{\rm TeV}$ are preferred to make the theory accessible to explain the excesses.
\item Eqs.~(\ref{Simplify-1}) and (\ref{New-mass-matrix}) indicate that $\kappa$ and $v_s$ have no direct impacts on the observables involved in this study except that they enter the scalar potential of the singlet field and thus affect the electroweak symmetry breaking. Consequently, only the naturalness in realizing the symmetry breaking determines their posterior PDFs~\cite{Cao:2018iyk}.
\end{itemize}

\begin{figure*}[t]
		\centering
		\resizebox{1.0\textwidth}{!}{
        \includegraphics{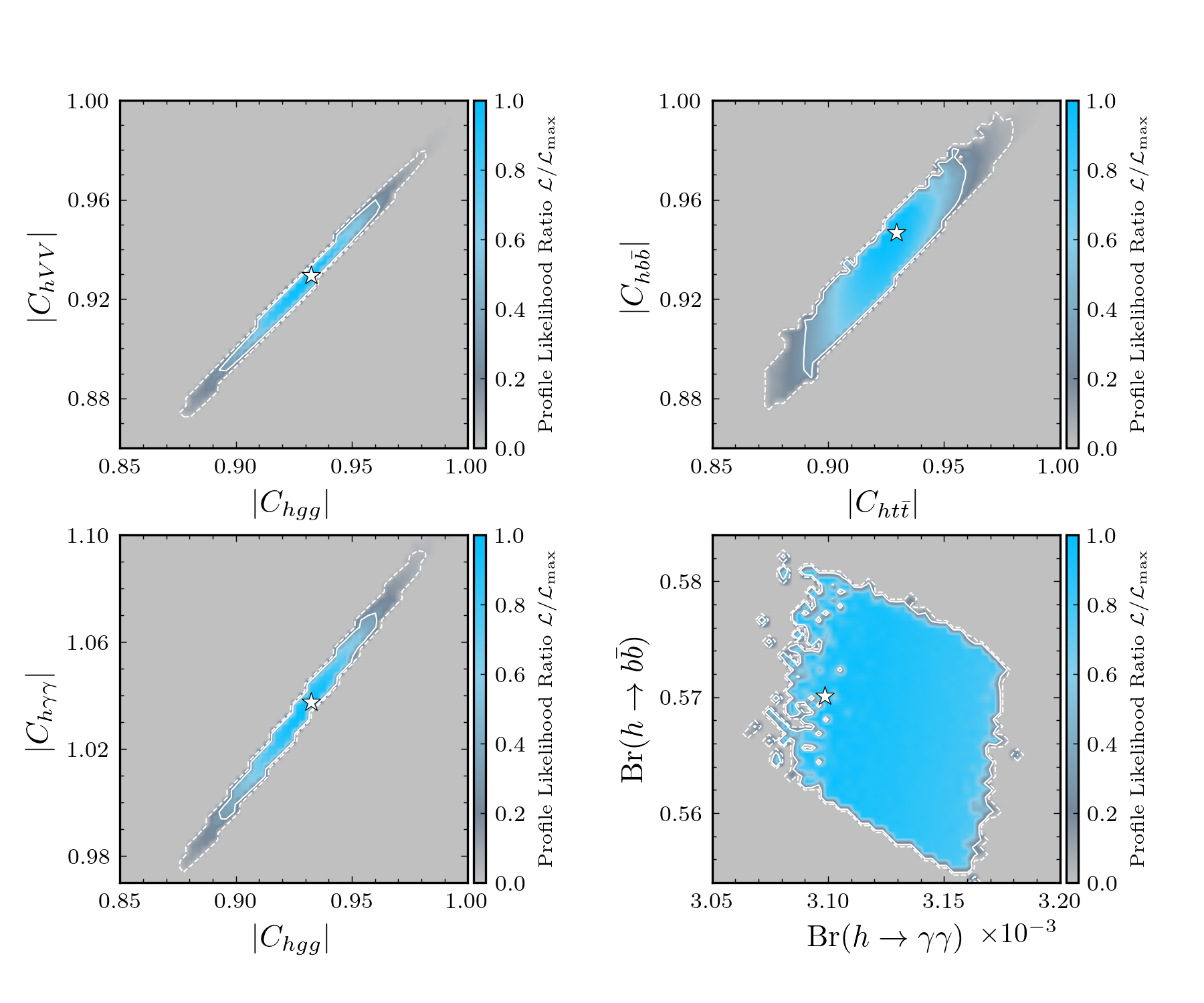}
        }

\vspace{-0.7cm}

\caption{Same as Fig.~\ref{Fig4}, but projected onto $|C_{h VV}| - |C_{h gg}|$, $|C_{h b\bar{b}}| - |C_{h t\bar{t}}|$, $|C_{h \gamma\gamma}| - |C_{h gg}|$, and $Br(h \to b\bar{b}) - Br(h \to \gamma\gamma)$ planes.   \label{Fig5}}
\end{figure*}

The scattering plots also reveal that the samples are hardly excluded by the LHC searches for electroweakinos. Specifically, the strategy of the LHC experiments in searching for SUSY relied on the mass splitting between the heavy sparticle produced at the LHC and the DM. It was categorized into the compressed spectrum case (Strategy I) and the significant mass splitting situation (Strategy II).  Strategy I could test the bino-wino coannihilation samples by the wino-pair productions and Strategy II by the higgsino-pair productions. However, the former way was less efficient in this study because there were no restrictions on the coannihilation case if $m_{\tilde{\chi}_1^0} \gtrsim 220~{\rm GeV}$, as shown in Fig. 16 of Ref.~\cite{ATLAS:2021moa}. The latter way was also inefficient because both the higgsinos and the DM were heavy after comparing the right panel in Fig.~\ref{Fig3} with Fig. 12 in Ref.~\cite{ATLAS:2021yqv}. Numerically speaking, we chose some points expected to have remarkable signals at the LHC and simulated the electroweakino productions with the package \texttt{CheckMATE-2.0.26}~\cite{Drees:2013wra,Dercks:2016npn, Kim:2015wza}. We concluded that the acquired R-values\footnote{The $R$-value is defined by $R \equiv max\{S_i/S_{i,obs}^{95}\}$, where $S_i$ denotes the simulated event number of the $i$th signal region in the analyses listed in Tables 1 of Ref.~\cite{Cao:2022ovk} and the research in Ref.~\cite{ATLAS:2021yqv}. $S_{i,obs}^{95}$ represents its $95\%$ confidence level upper limit. $R < 1$ implies that the sample is consistent with the LHC search for SUSY.} were always less than 0.4. The fundamental reason for the weak restrictions is that explaining the excesses heavily relies on the Higgs mixings instead of sizable supersymmetric contributions to $C_{h_s g g}$ and $C_{h_s \gamma \gamma}$, which need the presence of light sparticles. The GNMSSM predicts broad parameter space to tune these mixings freely.

\begin{figure*}[t]
		\centering
		\resizebox{1.0\textwidth}{!}{
        \includegraphics{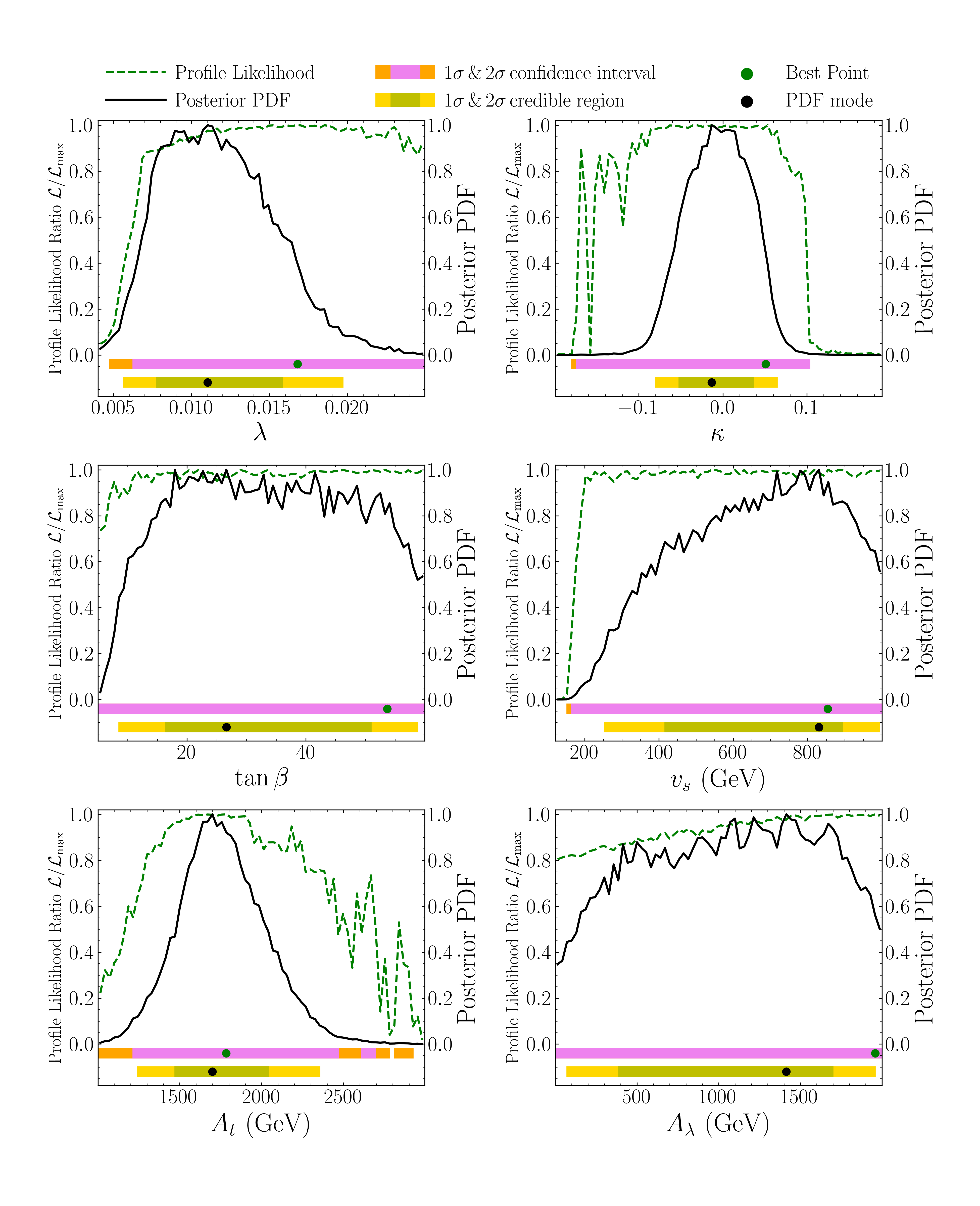}
        }

\vspace{-1.2cm}

\caption{Same as Fig.~\ref{Fig1}, but for the $\tilde{S}$-dominated DM case and $\chi^2_{\gamma \gamma + b\bar{b}} \simeq 0$ for the best point.  \label{Fig6}}
\end{figure*}

\begin{figure*}[thpb]
		\centering
		\resizebox{1.0\textwidth}{!}{
        \includegraphics{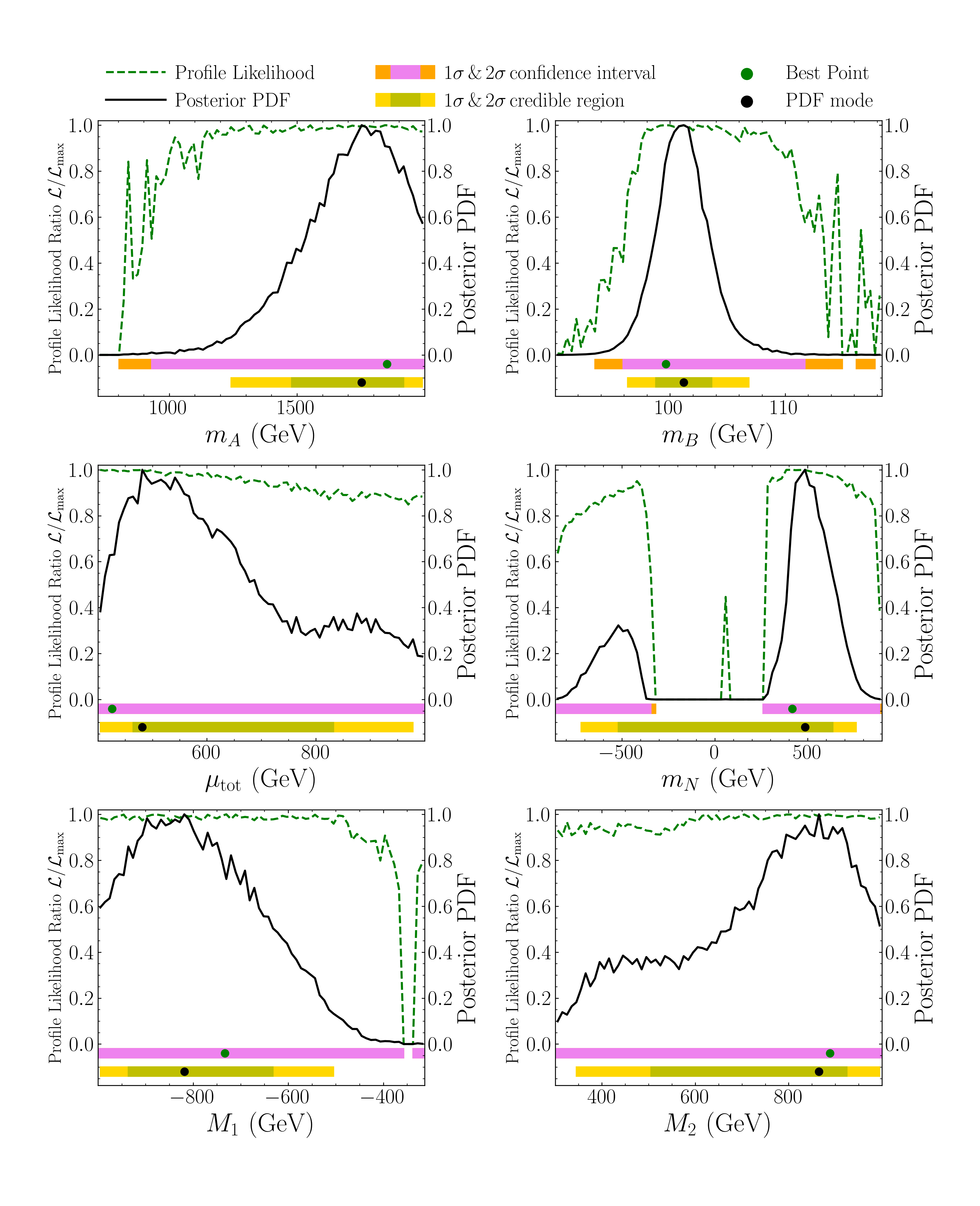}
        }

\vspace{-1.2cm}

\caption{Same as Fig.~\ref{Fig2}, but for the $\tilde{S}$-dominated DM case.  \label{Fig7}}
\end{figure*}

\begin{figure*}[thpb]
		\centering
		\resizebox{1.0\textwidth}{!}{
        \includegraphics{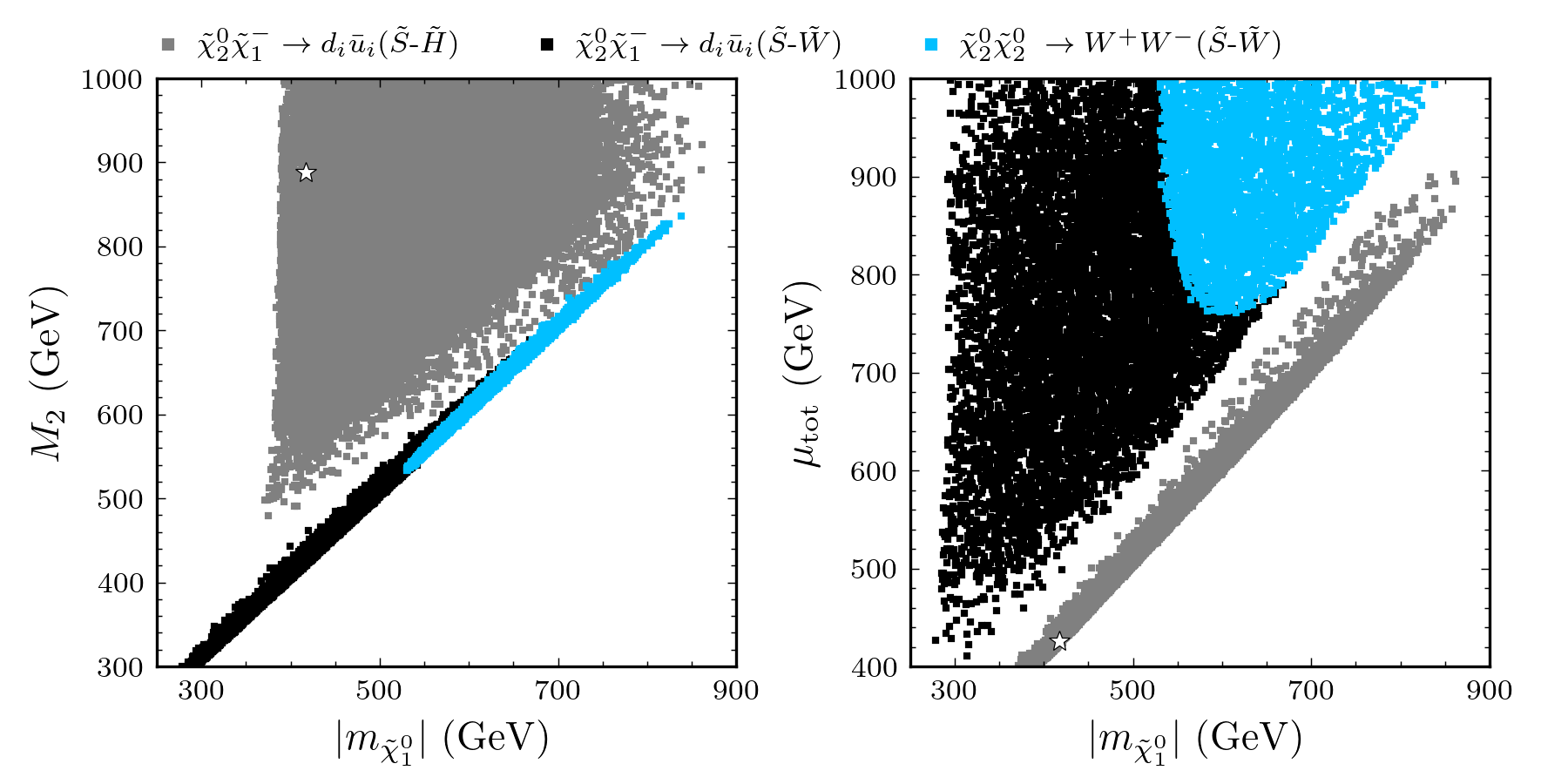}
        }

\vspace{-0.45cm}

\caption{Same as Fig.~\ref{Fig3}, but for the $\tilde{S}$-dominated DM case. It shows that the DM may co-annihilated with higgsino-like (in most cases) or wino-like electroweakinos to acquire the measured density. The colors distinguish the dominant annihilation channels.  \label{Fig8}}
\end{figure*}

\begin{figure*}[thpb]
		\centering
		\resizebox{1.0\textwidth}{!}{
        \includegraphics{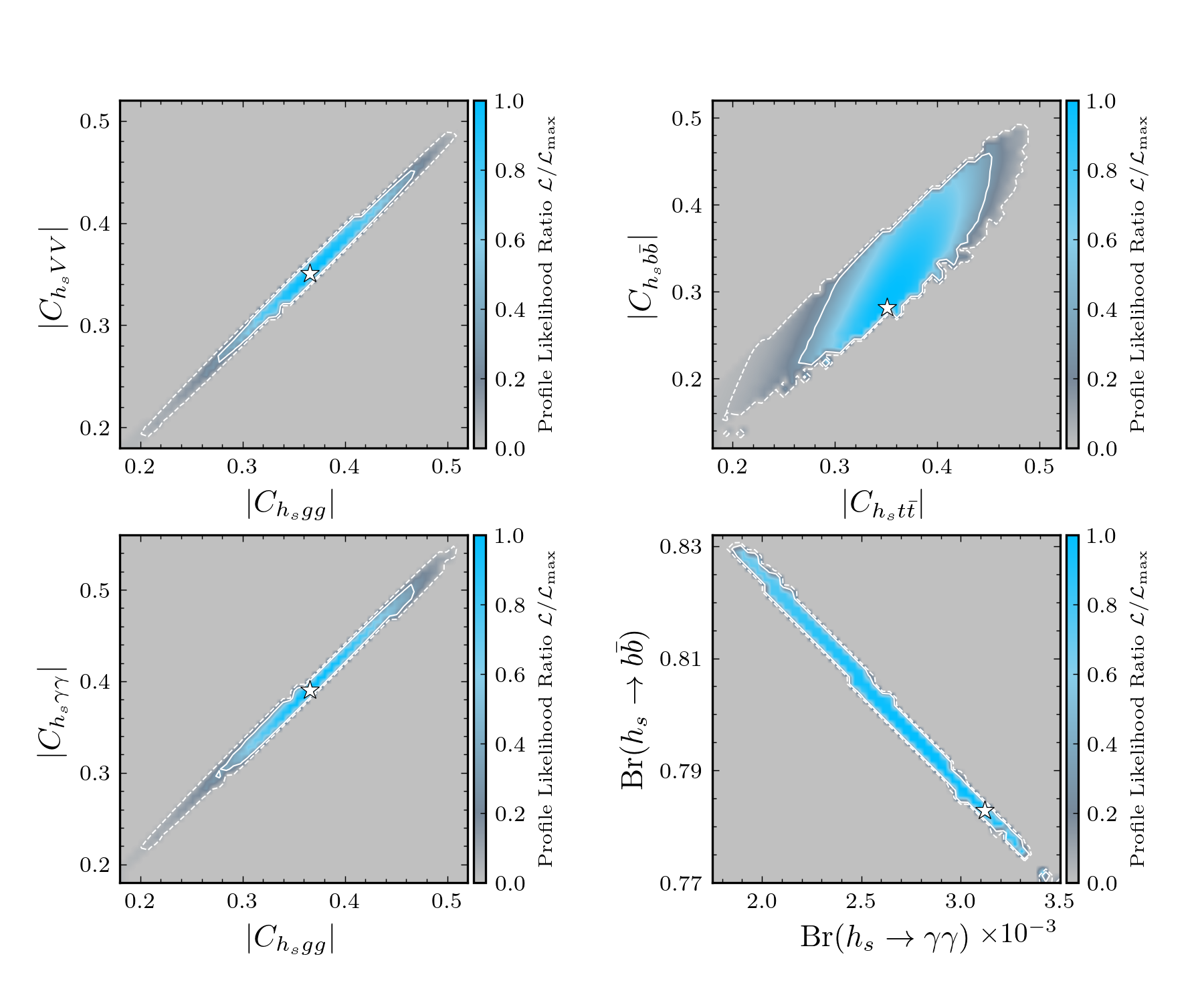}
        }

\vspace{-0.85cm}

\caption{Same as Fig.~\ref{Fig4}, but for the $\tilde{S}$-dominated DM case with $\chi^2_{\gamma \gamma + b \bar{b}}\simeq 0$ for the best point.   \label{Fig9}}
\end{figure*}

\begin{figure*}[thpb]
		\centering
		\resizebox{1.0\textwidth}{!}{
        \includegraphics{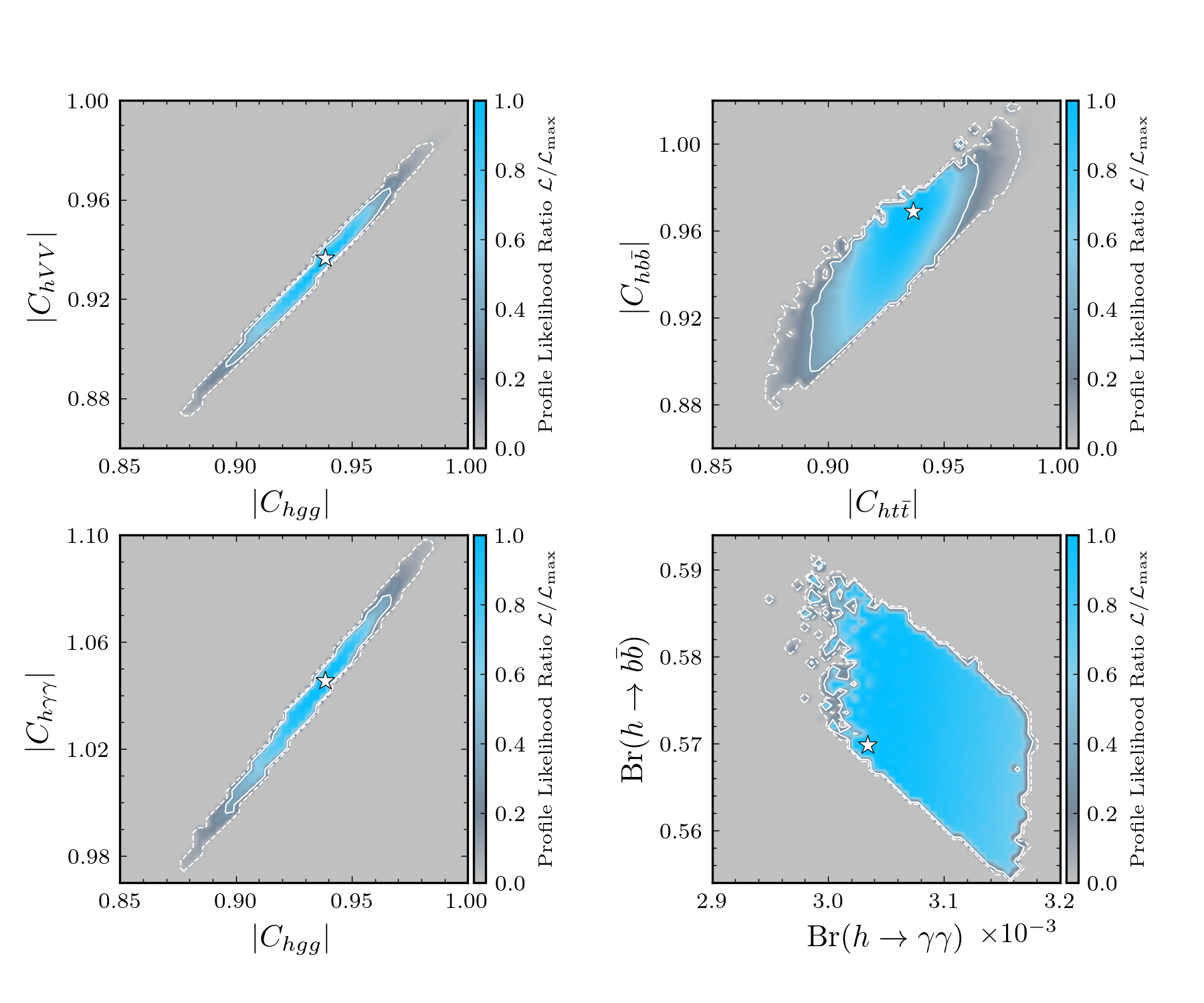}
        }

\vspace{-0.8cm}

\caption{Same as Fig.~\ref{Fig5}, but for the $\tilde{S}$-dominated DM case. \label{Fig10}}
\end{figure*}

Next, we studied the properties of the Higgs bosons. In Fig.~\ref{Fig4}, we plotted the two-dimensional PL maps on the $|C_{h_s VV}| - |C_{h_s gg}|$, $|C_{h_s b\bar{b}}| - |C_{h_s t\bar{t}}|$, $|C_{h_s \gamma\gamma}| - |C_{h_s gg}|$, and $Br(h_s\to b\bar{b}) - Br(h_s\to \gamma\gamma)$ planes. This figure reveals $|C_{h_s V V}| \simeq |C_{h_s t \bar{t}}|$, that $|C_{h_s g g}|$ and $|C_{h_s \gamma \gamma}|$ are slightly larger than $|C_{h_s t \bar{t}}|$, and that explaining the excesses at the $2\sigma$ level prefers the region characterized by $ 0.20 \lesssim |C_{h_s t \bar{t}}| \lesssim 0.50$, $ 0.20 \lesssim |C_{h_s b \bar{b}}| \lesssim 0.50$ with the correlation of $ |C_{h_s t \bar{t}}|/ |C_{h_s b \bar{b}}| \simeq 1.1$, $1.85 \times 10^{-3} \lesssim {\rm Br}_{\rm SUSY} (h_s \to \gamma \gamma) \lesssim 2.40 \times 10^{-3}$, and $ 81\% \lesssim {\rm Br}_{\rm SUSY} (h_s \to b \bar{b}) \lesssim 83 \%$. One can understand these features by the formulae of $\mu_{\gamma \gamma}$ and $\mu_{b \bar{b}}$ in Sec.~\ref{Section-excess}. Besides, the best point predicts $|C_{h_s t \bar{t}}| \simeq 0.37$, $|C_{h_s b \bar{b}}| \simeq 0.34$, ${\rm Br}_{\rm SUSY} (h_s \to \gamma \gamma) \simeq 2.4 \times 10^{-3}$, and ${\rm Br}_{\rm SUSY} (h_s \to b \bar{b}) \simeq 81\%$, which significantly deviate from the expectations in Sec.~\ref{Section-excess}. The fundamental reason is it predicts $\mu_{\gamma \gamma}= 0.206$ and $\mu_{b \bar{b}} = 0.135$, sizeably away from their experimental central values. By contrast, we will show in Fig.~\ref{Fig9} the best point for the $\tilde{S}$-dominated DM case, which predicts $\chi^2_{\gamma \gamma + b \bar{b}} \simeq 0$ and its other properties consistent with the expectations. We also studied the couplings of the SM-like Higgs boson in Fig.~\ref{Fig5}. This figure shows that the normalized couplings $|C_{h V V}|$, $|C_{h t \bar{t}}|$, $|C_{h b \bar{b}}|$, $|C_{h g g}|$, and $|C_{h \gamma \gamma}|$ are centered around 0.93, 0.93, 0.95, 0.93, and 1.04, respectively, in interpreting the excesses at the $2\sigma$ level. They agree with the SM predictions within $10\%$ uncertainties. ${\rm Br}(h \to b \bar{b})$ varies from $55.5 \%$ to $58.2 \%$, coinciding with its SM prediction of $(57.7 \pm 1.8)\%$~\cite{LHCHiggsCrossSectionWorkingGroup:2013rie}. In addition, although $Br (h \to \gamma \gamma)$  changes from $3.07 \times 10^{-3}$ to $3.16 \times 10^{-3}$, significantly larger than its SM prediction of $(2.28 \pm 0.11) \times 10^{-3}$~\cite{LHCHiggsCrossSectionWorkingGroup:2013rie}, the diphoton signal of $h$ is comparable with its SM prediction.

We verified that the mass of the charged Higgs boson varied from 930~{\rm GeV} to 3~{\rm TeV}, where the lower bound came from the restrictions of the LHC searches for extra Higgs bosons and the upper bound relied on the explored parameter space.

\subsubsection{Singlino-dominated DM case}

We studied the $\tilde{S}$-dominated DM case similarly. We showed the distributions of various parameters in Figs.~\ref{Fig6},~\ref{Fig7}, and~\ref{Fig8} and illustrated the Higgs properties in Figs.~\ref{Fig9} and~\ref{Fig10}. We learned the following differences after comparing these figures with their corresponding ones for the $\tilde{B}$-dominated DM case:
\begin{itemize}
\item The $\tilde{S}$-dominated DM achieved the measured abundance mainly by co-annihilating with higgsino-like (in most cases) or wino-like electroweakinos. Since the cross sections for the scattering of the $\tilde{S}$-dominated DM with nucleons were different from those for the $\tilde{B}$-dominated DM, as indicated by Eq. (2.30) of Ref.~\cite{Cao:2021ljw}, the LZ experiment allowed a moderately small $\mu_{tot}$ even in the singlino-higgsino co-annihilation case. Consequently, the posterior PDF preferred the spectrum pattern characterized by relatively small $|m_N|$ and $\mu_{tot}$ together with larger $|M_1|$ and $M_2$, as shown in Fig.~\ref{Fig7}. It also preferred a larger $\lambda$ than the prediction of the $\tilde{B}$-dominated DM case and subsequently a larger $V_{h_s}^{\rm NSM}$, as indicated in Eqs.~(\ref{Approximation-relations}-\ref{Approximation-relations-1}). This feature was crucial for the case to acquire the experimental values of $\mu_{\gamma \gamma}$ and $\mu_{b\bar{b}}$. In addition, given that the $t$-channel $h_s$-mediated contribution to the SI DM-nucleon scattering might be crucial with its amplitude proportional to $\kappa V_{h_s}^{\rm SM}$ in the leading-order approximation~\cite{Cao:2021ljw}, the LZ results preferred a small $\kappa$. Fig.~\ref{Fig6} exhibits this feature.

    Notably, the $\tilde{S}$-dominated DM can permanently annihilate into the $h_s h_s$ state. It may also annihilate into the $h_s A_s$ state if the kinematics is accessible.  However, these annihilation channels are never dominant since $\kappa$ is not significant~\cite{Cao:2021ljw}. Besides, the DM may achieve the meausred density by the $A_s$-meidated resonant annihilation. This case contributes to the total Bayesain evidence by about $0.2\%$, very small because it needs the tuning of $m_N$ and $m_{A_s}$ to satisfy $m_N \simeq m_{A_s}/2$.

\begin{figure*}[t]
		\centering
		\resizebox{1.0\textwidth}{!}{
        \includegraphics{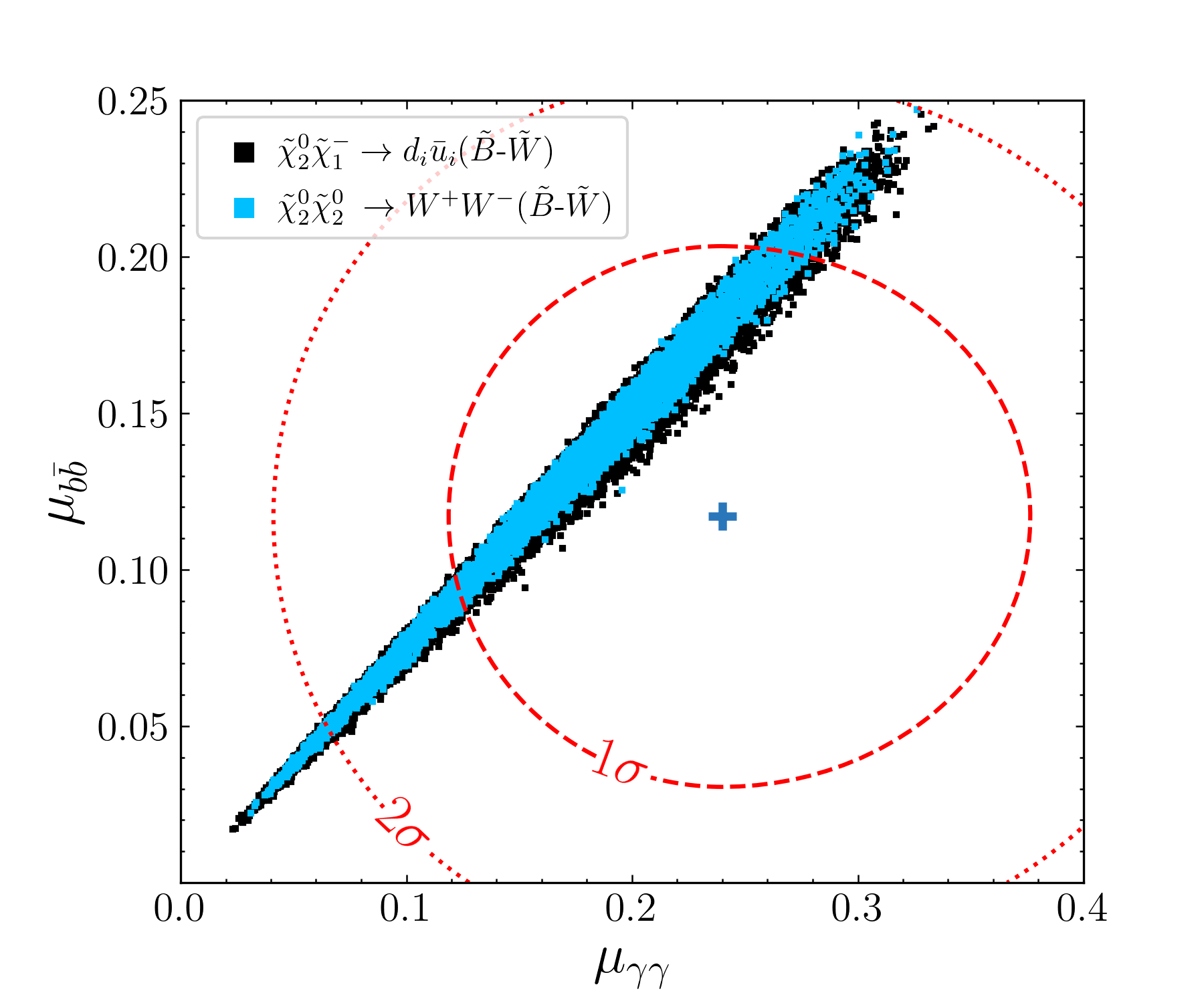}
        \includegraphics{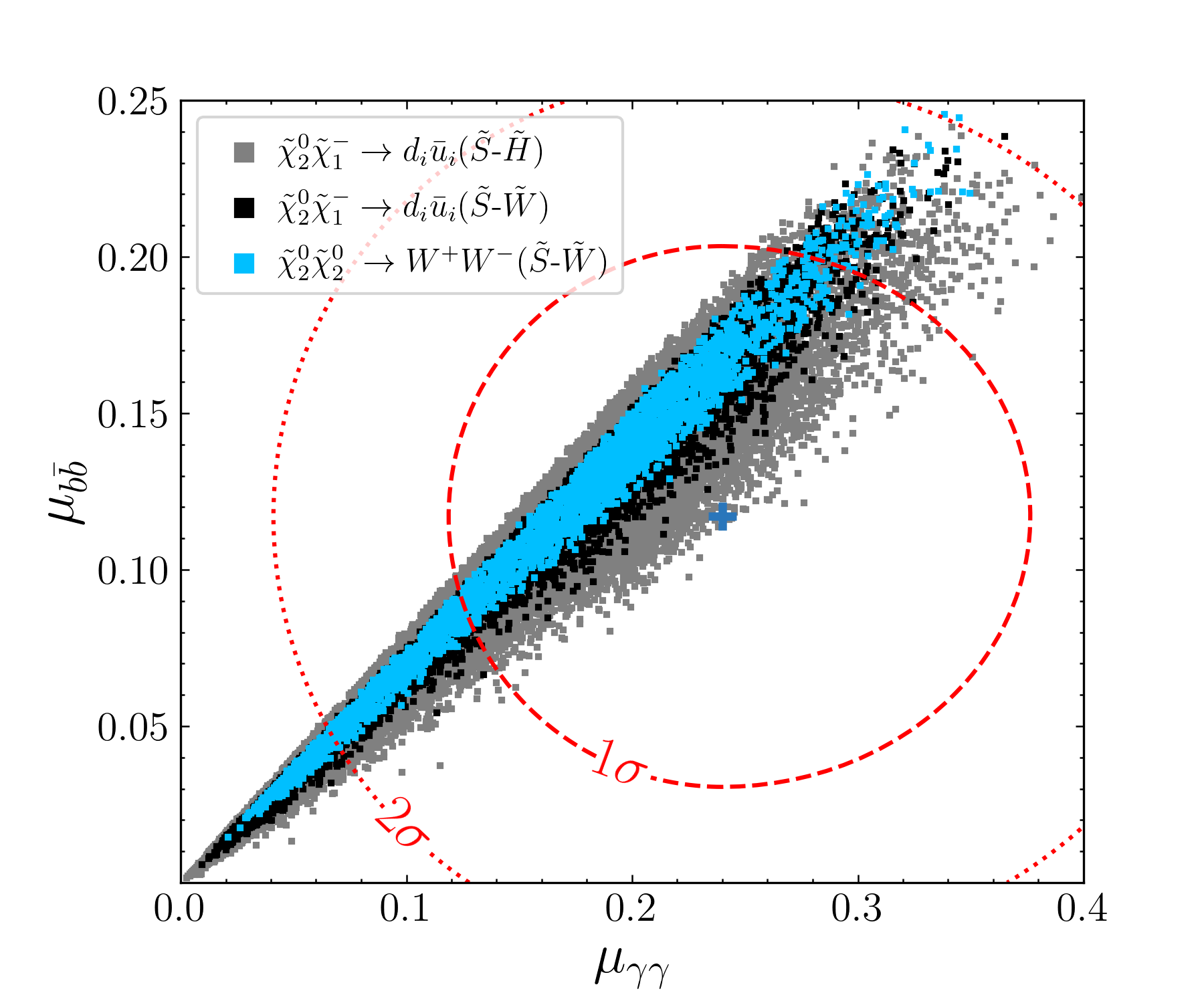}
        }

\vspace{-0.3cm}

\caption{Projection of the samples surviving the experimental restrictions onto the $\mu_{b\bar{b}}- \mu_{\gamma\gamma}$ planes. The left and right panels depict the pieces predicting $\tilde{B}$- and $\tilde{S}$-dominated DMs, respectively. The black and blue points represent the samples with $\tilde{\chi}_2^0 \tilde{\chi}_1^- \to d_i \bar{u}_i$ (i=1,2,3 denote the quark generations) and $\tilde{\chi}_2^0 \tilde{\chi}_2^0 \to W^+ W^-$ acting as the dominant annihilation channels in the bino-wino co-annihilation  case (left panel) and the singlino-wino co-annihilation case (right panel), and the grey ones correspond to $\tilde{\chi}_2^0 \tilde{\chi}_1^- \to d_i \bar{u}_i$ (i=1,2,3) as the dominant annihilation channel in the singlino-higgsino co-annihilation case.
Samples in the region enclosed by the red dashed and red dotted lines satisfy $\chi^2_{\gamma\gamma + b\bar{b}} \leq 2.30$ and $\chi^2_{\gamma\gamma + b\bar{b}} \leq 6.18$, respectively, indicating that they can explain the diphoton and $b\bar{b}$ excesses at the $1\sigma$ and $2\sigma$ levels. \label{Fig11}}
\end{figure*}

\item Since the $\tilde{S}$-dominated DM had very weak couplings to the other sparticles, the heavy ones, except for the next-to-lightest sparticle (NLSP), were unlikely to directly decay into the DM. As a result, their decay chains were usually lengthened, which complicated the SUSY search. Concerning the co-annihilation cases, however, one might simplify the situation. Specifically, the particle X in the decay ${\rm NLSP} \to {\rm DM} + {\rm X}$ was too soft to be detected by Strategy II, and thus, the NLSP behavior as the missing momentum at the LHC, the same as the DM signal. Consequently, the LHC search for SUSY had no exclusion capability on the samples after comparing the sparticle spectrum in Fig.~\ref{Fig8} with the last two panels in Fig. 14 of Ref.~\cite{ATLAS:2021yqv}, given that the DM and its co-annihilation partners were massive, i.e., $m_{\tilde{\chi}_{1,2}^0}$, $m_{\tilde{\chi}_1^\pm} \gtrsim 300~{\rm GeV}$ with $m_{\tilde{\chi}_2^0} \simeq m_{\tilde{\chi}_1^\pm} \simeq m_{\tilde{\chi}_1^0}$ for the singlino-wino co-annihilation case and  $m_{\tilde{\chi}_{1,2,3}^0}$, $m_{\tilde{\chi}_1^\pm} \gtrsim 400~{\rm GeV}$ with $m_{\tilde{\chi}_2^0} \simeq m_{\tilde{\chi}_3^0} \simeq m_{\tilde{\chi}_1^\pm} \simeq m_{\tilde{\chi}_1^0}$ for the singlino-higgsino co-annihilation case. We added that the Strategy I also failed to exclude the compressed mass spectrum of the $\tilde{S}$-dominated DM case, again because the DM was massive. Furthermore, we studied some samples expected to leave remarkable signals at the LHC. Our simulation with the package \texttt{CheckMATE-2.0.26} revealed that the R-values were usually less than 0.5.
\item Comparing Figs.~\ref{Fig4} and \ref{Fig9} indicated that the $2\sigma$ confidence intervals for the $\tilde{S}$-dominated DM case were slightly larger than those for the $\tilde{B}$-dominated DM case, and the $\chi^2_{\gamma \gamma + b \bar{b}}$ for the best point of the former case was significantly smaller than that of the latter case. These observations revealed that the $\tilde{S}$-dominated DM case was more suited to explain the excesses. One primary reason was that the $\tilde{S}$-dominated DM case allowed a larger $\lambda$ and a smaller $\mu_{tot}$. It could predict a significant $V_{h_s}^{\rm NSM}$ and slightly enhance the supersymmetric contribution to $C_{h_s \gamma \gamma}$.  Both could mitigate the substantial suppression of $C_{h_s b \bar{b}}$ and weaken the correlation between $\mu_{\gamma \gamma}$ and $\mu_{b \bar{b}}$. They make the case more accessible to explain the excess.
\end{itemize}

\subsubsection{Status of the excesses}

\begin{table}[t]
%\begin{sidewaystable}
\centering
\resizebox{1\textwidth}{!}
 {
\begin{tabular}{lrlr|lrlr}
\hline \hline
\multicolumn{4}{c|}{\bf Benchmark Point P1}                                                                                                								& \multicolumn{4}{c}{\bf Benchmark Point P2}
 \\ \hline
$\lambda$             					& 0.012& $m_{h_s}$                			& 96.0~GeV& $\lambda$             				& 0.017& $m_{h_s}$                		& 96.1~GeV\\
$\kappa$              					&  0.133&  $m_{h}$                				& 125.1~GeV& $\kappa$              				& 0.051& $m_{h}$              			& 125.5~GeV\\
$\tan{\beta}$         					& 56.41&  $m_{H}$                				& 2183~GeV& $\tan{\beta}$         				& 53.71& $m_{H}$                  			& 2232~GeV\\
$v_s$                						& 520.5~GeV&  $m_{A_s}$             		 		& 797.8~GeV& $v_s$                 				& 853.9~GeV& $m_{A_s}$                 		& 800.0~GeV\\
$\mu_{\rm tot}$						& 576.0~GeV& $m_{A_H}$                			& 2183~GeV& $\mu_{\rm tot}$ 				& 425.7~GeV& $m_{A_H}$                		& 2232~GeV\\
$M_1$								& -294.0~GeV& $m_{H^\pm}$					& 2219~GeV& $M_1$							& -733.0~GeV& $m_{H^\pm}$				&2257~GeV\\
$M_2$ 								& 305.2~GeV& $m_{\tilde{\chi}_1^0}$   			& -293.9~GeV& $M_2$                  				& 888.3~GeV& $m_{\tilde{\chi}_1^0}$   		& 417.2~GeV\\
$A_t$                 						& 1522 ~GeV& $m_{\tilde{\chi}_2^0}$   			& 317.3~GeV& $A_t$	   						& 1784~GeV& $m_{\tilde{\chi}_2^0}$   		& 432.6~GeV\\
$A_\lambda$& 1589 ~GeV& $m_{\tilde{\chi}_3^0}$			& -416.1~GeV& $A_\lambda$& 1959~GeV& $m_{\tilde{\chi}_3^0}$		& -436.8~GeV\\
$m_A$& 1683 ~GeV& $m_{\tilde{\chi}_4^0}$ 			& -595.3~GeV& $m_A$& 1853 ~GeV& $m_{\tilde{\chi}_4^0}$ 		& -741.3~GeV\\
$m_B$& 105.82 ~GeV&$m_{\tilde{\chi}_5^0}$ 			& 601.9~GeV& $m_B$& 99.65 ~GeV&$m_{\tilde{\chi}_5^0}$ 		& 930.7~GeV\\
 $m_N$& -416.5 ~GeV& $m_{\tilde{\chi}_1^\pm}$	 		& 317.5~GeV& $m_N$& 417.6 ~GeV& $m_{\tilde{\chi}_1^\pm}$	 	& 433.9~GeV\\
$\mu_{\gamma\gamma}$&  0.206& $m_{\tilde{\chi}_2^\pm}$			& 604.4~GeV& $\mu_{\gamma\gamma}$&  0.243& $m_{\tilde{\chi}_2^\pm}$		& 930.9~GeV\\
$\mu_{b\bar{b}}$& 0.134&$\sigma^{SI}_p$        & $6.75\times 10^{-47}{\rm ~cm^2}$& $\mu_{b\bar{b}}$& 0.116&$\sigma^{SI}_p$        &$1.07\times 10^{-46}{\rm ~cm^2}$\\
$\Omega h^2$& 0.113&$\sigma^{SD}_n$       &$1.55\times 10^{-42}{\rm ~~cm^2}$& $\Omega h^2$& 0.131&$\sigma^{SD}_n$       &$5.60\times 10^{-45}{\rm ~~cm^2}$\\
\hline
\multicolumn{2}{l}{$V_{h_s}^{S}, ~V_{h_s}^{SM}, ~V_{h_s}^{NSM}$}  & \multicolumn{2}{l|}{-0.929, ~0.369, ~$4.5\times 10^{-4}$}
& \multicolumn{2}{l}{$V_{h_s}^{S}, ~V_{h_s}^{SM}, ~V_{h_s}^{NSM}$}  & \multicolumn{2}{l}{-0.936, ~0.351, ~$1.3\times 10^{-3}$} \\
\multicolumn{2}{l}{$N_{11}, ~N_{12}, ~N_{13}, ~N_{14}, ~N_{15}$}   &\multicolumn{2}{l|}{~0.994, ~~0.006,  ~~0.097, ~~0.047,  ~-0.002}
& \multicolumn{2}{l}{$N_{11}, ~N_{12}, ~N_{13}, ~N_{14}, ~N_{15}$}   &\multicolumn{2}{l}{~0.004, ~~0.015, ~-0.092, ~~0.095, ~-0.991} \\
\multicolumn{2}{l}{$N_{21}, ~N_{22}, ~N_{23}, ~N_{24}, ~N_{25}$}   &\multicolumn{2}{l|}{-0.007,~~-0.978,~~~0.184,~~-0.100,~~-0.001}
& \multicolumn{2}{l}{$N_{21}, ~N_{22}, ~N_{23}, ~N_{24}, ~N_{25}$}   &\multicolumn{2}{l}{-0.026,~~-0.113,~~~0.700,~~-0.691,~~-0.133} \\
\multicolumn{2}{l}{$N_{31}, ~N_{32}, ~N_{33}, ~N_{34}, ~N_{35}$}   &\multicolumn{2}{l|}{-0.003,~~~0.001,~~~0.004,~~~0.007,~~-0.999}
& \multicolumn{2}{l}{$N_{31}, ~N_{32}, ~N_{33}, ~N_{34}, ~N_{35}$}   &\multicolumn{2}{l}{-0.101,~~-0.040,~~-0.704,~~-0.702, ~-0.002} \\
\multicolumn{2}{l}{$N_{41}, ~N_{42}, ~N_{43}, ~N_{44}, ~N_{45}$}   &\multicolumn{2}{l|}{-0.102,~~~0.060,~~~0.698,~~~0.706,~~~0.008}
& \multicolumn{2}{l}{$N_{41}, ~N_{42}, ~N_{43}, ~N_{44}, ~N_{45}$}   &\multicolumn{2}{l}{~0.995,~~-0.004,~~-0.052,~~-0.090,~~-0.000} \\
\multicolumn{2}{l}{$N_{51}, ~N_{52}, ~N_{53}, ~N_{54}, ~N_{55}$}   &\multicolumn{2}{l|}{~0.035,~~-0.201,~~-0.685,~~~0.699,~~~0.001}
& \multicolumn{2}{l}{$N_{51}, ~N_{52}, ~N_{53}, ~N_{54}, ~N_{55}$}   &\multicolumn{2}{l}{~0.003,~~-0.993,~~-0.053,~~~0.109,~~~0.000} \\
\multicolumn{2}{l}{$C_{h_s gg}, ~C_{h_s VV}, ~C_{h_s \gamma\gamma}, ~C_{h_s t \bar{t}}, ~C_{h_s b\bar{b}}$} 	& \multicolumn{2}{l|}{~0.383, ~~0.369,~~~0.411,~~~0.369,~~0.344}
&\multicolumn{2}{l}{$C_{h_s gg}, ~C_{h_s VV}, ~C_{h_s \gamma\gamma}, ~C_{h_s t \bar{t}}, ~C_{h_s b\bar{b}}$} 	& \multicolumn{2}{l}{~0.366,~~~0.351,~~~0.391,~~~0.351,~~0.281} \\
 \multicolumn{2}{l}{$C_{h gg}, ~~C_{h VV}, ~~C_{h \gamma\gamma}, ~~C_{h t \bar{t}}, ~~C_{h b\bar{b}}$} 			& \multicolumn{2}{l|}{~0.932,~~~0.929, ~~1.037,~~~0.929,~~~0.947}
& \multicolumn{2}{l}{$C_{h gg}, ~~C_{h VV}, ~~C_{h \gamma\gamma}, ~~C_{h t \bar{t}}, ~~C_{h b\bar{b}}$} 			& \multicolumn{2}{l}{~0.938,~~~0.936, ~~1.046,~~~0.936,~~~0.969} \\
\hline
\multicolumn{2}{l}{ Coannihilations }                         & \multicolumn{2}{l|}{Fractions [\%]} 								& \multicolumn{2}{l}{Coannihilations}                                       & \multicolumn{2}{l}{Fractions [\%]}                                      \\
\multicolumn{2}{l}{$\tilde{\chi}_2^0\tilde{\chi}_1^- \to d_i \bar{u}_i / Z W^- / \nu_{\ell_i} \ell_i^- / A_s W^- $} 						& \multicolumn{2}{l|}{29.3/6.5/10.2/1.7}
& \multicolumn{2}{l}{$\tilde{\chi}_2^0\tilde{\chi}_1^- \to d_i \bar{u}_i / \nu_{\ell_i} \ell_i^- / Z W^- / A_s W^- / h W^-  $} 	& \multicolumn{2}{l}{21.8/8.0/1.3/1.3/1.1}          \\
\multicolumn{2}{l}{$\tilde{\chi}_2^0 \tilde{\chi}_2^0 \to W^- W^+ $} 										& \multicolumn{2}{l|}{7.8}
& \multicolumn{2}{l}{$\tilde{\chi}_3^0\tilde{\chi}_1^- \to d_i \bar{u}_i / \nu_{\ell_i} \ell_i^- / Z W^- $} 	& \multicolumn{2}{l}{14.8/5.5/1.2}   \\  \hline
\multicolumn{2}{l}{ Decays }                         & \multicolumn{2}{l|}{Branching ratios [\%]} 								& \multicolumn{2}{l}{Decays}                                       & \multicolumn{2}{l}{Branching ratios [\%]}
\\
\multicolumn{2}{l}{$\tilde{\chi}^0_2 \to \tilde{\chi}^0_1 Z^\ast$}      & \multicolumn{2}{l|}{~100}
&\multicolumn{2}{l}{$\tilde{\chi}^0_2 \to \tilde{\chi}^0_1 Z^\ast$}      & \multicolumn{2}{l}{~100} \\

\multicolumn{2}{l}{$\tilde{\chi}^0_3 \to \tilde{\chi}^\pm_1 W^\mp / \tilde{\chi}^0_1 h_s$}      & \multicolumn{2}{l|}{~86.8/10.2}
&\multicolumn{2}{l}{$\tilde{\chi}^0_3 \to \tilde{\chi}^0_1 Z^\ast$}      & \multicolumn{2}{l}{~98.1} \\

\multicolumn{2}{l}{$\tilde{\chi}^0_4 \to \tilde{\chi}^\pm_1 W^\mp / \tilde{\chi}^0_2 Z / \tilde{\chi}^0_1 h$}      & \multicolumn{2}{l|}{~61.8/26.7/7.0}
&\multicolumn{2}{l}{$\tilde{\chi}^0_4 \to \tilde{\chi}^\pm_1 W^\mp / \tilde{\chi}^0_2 Z / \tilde{\chi}^0_3 h$}      & \multicolumn{2}{l}{~50.6/22.2/20.4} \\
\multicolumn{2}{l}{$\tilde{\chi}^0_5 \to \tilde{\chi}^\pm_1 W^\mp / \tilde{\chi}^0_2 h / \tilde{\chi}^0_1 Z$}      & \multicolumn{2}{l|}{~63.1/21.1/8.3}
&\multicolumn{2}{l}{$\tilde{\chi}^0_5 \to \tilde{\chi}^\pm_1 W^\mp / \tilde{\chi}^0_3 Z / \tilde{\chi}^0_2 h$}      & \multicolumn{2}{l}{~51.4/21.9/18.3} \\
\multicolumn{2}{l}{$\tilde{\chi}^+_1 \to \tilde{\chi}^0_1 (W^+)^\ast$}      & \multicolumn{2}{l|}{~100}
&\multicolumn{2}{l}{$\tilde{\chi}^+_1 \to \tilde{\chi}^0_1 (W^+)^\ast$}      & \multicolumn{2}{l}{~100}  \\
\multicolumn{2}{l}{$\tilde{\chi}^+_2 \to \tilde{\chi}^0_2 W^+ / \tilde{\chi}^+_1 Z / \tilde{\chi}^+_1 h / \tilde{\chi}^0_1 W^+$}      & \multicolumn{2}{l|}{~32.0/30.5/23.0/10.4}
&\multicolumn{2}{l}{$\tilde{\chi}^+_2 \to \tilde{\chi}^0_3 W^+ / \tilde{\chi}^+_1 Z / \tilde{\chi}^0_2 W^+ / \tilde{\chi}^+_1 h$}      & \multicolumn{2}{l}{~25.4/25.1/25.0/21.2}
\\
\multicolumn{2}{l}{$h_s \to b \bar{b} / \tau^+ \tau^- / gg / c \bar{c} / \gamma \gamma$}      & \multicolumn{2}{l|}{~81.1/9.0/4.8/4.4/0.0024}
&\multicolumn{2}{l}{$h_s \to b \bar{b} / \tau^+ \tau^- / gg / c \bar{c} / \gamma \gamma$}      & \multicolumn{2}{l}{~78.3/8.7/6.3/5.8/0.0031}
\\
\multicolumn{2}{l}{$h \to b \bar{b} / W W^\ast / \tau^+ \tau^- / gg / \gamma \gamma$}      & \multicolumn{2}{l|}{~57.0/26.4/6.6/4.6/0.0031}
&\multicolumn{2}{l}{$h \to b \bar{b} / W W^\ast / \tau^+ \tau^- / gg / \gamma \gamma$}      & \multicolumn{2}{l}{~57.0/26.6/6.6/4.5/0.0030}
\\
\multicolumn{2}{l}{$H \to b \bar{b} / \tau^+ \tau^- / \tilde{\chi}^+_1 \tilde{\chi}^-_2 / \tilde{\chi}^+_2 \tilde{\chi}^-_1$}      & \multicolumn{2}{l|}{~58.7/14.7/7.4/7.4}
&\multicolumn{2}{l}{$H \to b \bar{b} / \tau^+ \tau^- / \tilde{\chi}^+_1 \tilde{\chi}^-_2 / \tilde{\chi}^+_2 \tilde{\chi}^-_1$}      & \multicolumn{2}{l}{~63.6/14.6/6.3/6.3}
\\
\multicolumn{2}{l}{$A_H \to b \bar{b} / \tau^+ \tau^- $}      & \multicolumn{2}{l|}{~80.2/12.2}
&\multicolumn{2}{l}{$A_H \to b \bar{b} / \tau^+ \tau^- $}      & \multicolumn{2}{l}{~86.6/13.2}
\\
\multicolumn{2}{l}{$H^+ \to t \bar{b} / \tau^+ \nu_{\tau} / \tilde{\chi}^0_2 \tilde{\chi}^+_2 / \tilde{\chi}^0_5 \tilde{\chi}^+_1 / \tilde{\chi}^0_4 \tilde{\chi}^+_1 $}      & \multicolumn{2}{l|}{~57.6/16.1/8.1/7.6/7.2}
&\multicolumn{2}{l}{$H^+ \to t \bar{b} / \tau^+ \nu_{\tau} / \tilde{\chi}^0_2 \tilde{\chi}^+_2 / \tilde{\chi}^0_5 \tilde{\chi}^+_1 / \tilde{\chi}^0_3 \tilde{\chi}^+_2$}      & \multicolumn{2}{l}{~62.5/15.9/6.3/6.3/6.2}
\\ \hline
\multicolumn{2}{l}{$R$ value: 0.17}  & \multicolumn{2}{l|}{Signal Region: SR-WZoff-high-nja in Ref.~\cite{ATLAS:2021moa}}
& \multicolumn{2}{l}{$R$ value: 0.35}  & \multicolumn{2}{l}{Signal Region: SR-4Q-VV in Ref.~\cite{ATLAS:2021yqv}} \\
\hline \hline
\end{tabular}}
\caption{\label{tab:2} Details of the best points for the $\tilde{B}$- and $\tilde{S}$-dominated DM cases, respectively. Both can explain the diphoton and $b\bar{b}$ excesses at the $1\sigma$ level and simultaneously be consistent with the other experimental restrictions. They achieved the measured abundance by co-annihilating with the wino- and higgsino-dominated electroweakinos, respectively. $d_i$, $u_i$, and $\ell_i$ appear in the annihilation final state and denote the $i$th generation of down-type quarks, up-type quarks, and leptons, respectively.}
\end{table}

We summarize the status of the excesses in the GNMSSM. In Fig.~\ref{Fig11}, we projected all the samples acquired by the scan and survived the experimental restrictions onto the $\mu_{b\bar{b}}- \mu_{\gamma\gamma}$ planes. The left and right panels are for the $\tilde{B}$- and $\tilde{S}$-dominated DM cases, respectively, and the colors distinguish the dominant annihilation channels. This figure reveals that $\mu_{\gamma \gamma}$ and $\mu_{b\bar{b}}$ can reach 0.39 and 0.25, respectively, and their ratio, expressed as
\begin{eqnarray}
\frac{\mu_{\gamma \gamma}}{\mu_{b \bar{b}}} \equiv \frac{|C_{h_s g g}|^2}{|C_{h_s V V}|^2} \times \frac{|C_{h_s \gamma \gamma}|^2}{|C_{h_s b \bar{b}}|^2} \simeq \frac{|C_{h_s \gamma \gamma}|^2}{|C_{h_s b \bar{b}}|^2},
\end{eqnarray}
varies from 1.2 to 1.7 for the $\tilde{B}$-dominated DM case and 1.2 to 2.5 for the $\tilde{S}$-dominated DM case. These conclusions indicate that the GNMSSM can easily explain the diphoton and $b\bar{b}$ excesses at the $1 \sigma$ level. In particular, the $\tilde{S}$-dominated DM case can predict the central values of the signal strengths for the excesses. We verified that allowing $A_\lambda$ to vary within a broader range than that in Table~\ref{tab:1} could enhance the maximum reach of the ratio and improve the GNMSSM's capability to explain the excesses.

To further illuminate the physics underlying the excesses, we presented in Table~\ref{tab:2} the details of two benchmark points, which corresponded to the best points of the two types of DM cases, respectively. Both explain the di-photon and $b\bar{b}$ excesses at the $1\sigma$ level and agree well with the other experimental restrictions. Particularly, their $R$ values are not much below 1, implying that they will be explored at the LHC by the electroweakino productions in future.

\section{Implications of the excesses \label{Section-implication}}

The GNMSSM interpretation of the excesses will be tested at future linear colliders, either by searching for $h_s$ via the process $e^+ e^- \to Z h_s$ and determining its properties or by precisely measuring
the couplings of $h$~\cite{Biekotter:2019mib}. It may also be explored by searching for the doublet-dominated Higgs bosons, $H$, $A_H$, and $H^\pm$, or the electroweakinos, assuming these particles are moderately light~\cite{Biekotter:2020cjs,Dutta:2023cig}. In this section, we briefly discuss the other signals of $h_s$ at the LHC.

In Ref.~\cite{CMS:2022tgk}, the CMS collaboration searched for the resonant production of a scalar $X$ by the channel $p p \to X \to h Y \to (\gamma \gamma) (b \bar{b})$ with an integrated luminosity of $138 {\rm fb}^{-1}$ at $\sqrt{s} = 13~{\rm TeV}$, where $Y$ denoted another scalar satisfying $m_Y < m_X - m_h$. They acquired the $95\%$ confidence level  upper limits on the $\gamma \gamma b \bar{b}$ signal rate, ranging from $0.04~{\rm fb}$ to $0.90~{\rm fb}$. Taking $X=H$ and $Y=h_s$ in the GNMSSM, the cross section of the $\gamma \gamma b \bar{b}$ signal is given by
\begin{eqnarray}
\sigma_{\gamma \gamma b \bar{b}} &=& (\sigma_{\rm SM}^{g g H} |C_{H g g}|^2 + \sigma_{\rm SM}^{b \bar{b} H} |C_{H b \bar{b}}|^2) \times Br(H \to h h_s) \times Br(h \to \gamma \gamma) \times Br(h_s \to b \bar{b}) \nonumber \\
& \simeq & (\sigma_{\rm SM}^{g g H} \times m_b^2/m_t^2 + \sigma_{\rm SM}^{b \bar{b} H} ) \times \tan^2 \beta \times Br(H \to h h_s) \times 0.228\% \times 76\%,
\end{eqnarray}
in the large $\tan \beta$ case, where $\sigma_{\rm SM}^{g g H}$ and $\sigma_{\rm SM}^{b \bar{b} H}$ are the SM prediction of the $H$ production rate via gluon-gluon and $b \bar{b}$ fusions, respectively, and $Br(h_s \to b \bar{b}) \simeq 76\%$ as discussed in Sec.~\ref{Section-excess}. If the decays of $H$  into sparticles are kinematically forbidden,  $H \to b \bar{b}$ will be the dominant decay channel and thus $Br(H \to h h_s) \simeq \Gamma ( H \to h h_s)/\Gamma (H \to b \bar{b})$.  Since $|V_H^{\rm NSM}| \simeq 1$, $|V_h^{\rm SM}| \simeq 1$, and  $|V_{h_s}^{\rm S}| \simeq 1$, the $H h h_s$ coupling normalized by a factor of $-m_Z^2/v$~\cite{Djouadi:2005gj} can be acquired from Eq. (A.15) of Ref.~\cite{Ellwanger:2009dp}. It is $\lambda_{H h h_s} \simeq \lambda (A_\lambda + m_N) v/m_Z^2 \simeq - m_A^2 V_{h_s}^{\rm NSM}/m_Z^2$, where we used the first approximation in Eq.~(\ref{Approximation-relations}) in the last step. Given $m_H \gg m_h$, we concluded that $Br(H \to h h_s) \simeq m_H^2 |V_{h_s}^{\rm NSM}|^2/(12 m_b^2 \tan^2 \beta)$ by the width formulae in Ref.~\cite{Djouadi:2005gj} and
\begin{eqnarray}
\sigma_{\gamma \gamma b \bar{b}} \simeq 7.1 \times 10^{-11} \times ( 5.8 \sigma_{\rm SM}^{g g H} + 10^4 \sigma_{\rm SM}^{b \bar{b} H} ) \times \left ( \frac{m_H}{m_b \tan \beta} \right )^2 \times \left ( \frac{V_{h_s}^{\rm NSM} \tan \beta}{0.07} \right )^2.
\end{eqnarray}
These formulae indicate that on the premise of explaining the diphoton and $b\bar{b}$ excesses, $\sigma_{\gamma \gamma b \bar{b}} \simeq 0.026/\tan^2 \beta~{\rm fb}$ for $m_H = 650~{\rm GeV}$ and $\sigma_{\gamma \gamma b \bar{b}} \simeq 0.014/\tan^2 \beta~{\rm fb}$ for $m_H = 800~{\rm GeV}$\footnote{In selecting these benchmark values of $m_H$, we do not consider the restrictions from the LHC searches for extra Higgs bosons by $\tau \bar{\tau}$ signal, which have set a bound of $m_H \gtrsim 930~{\rm GeV}$ in this study.}. These cross sections are consistent with corresponding experimental bounds of $0.36~{\rm fb}$ and $0.31~{\rm fb}$, respectively. Alternatively, the formulae imply that the GNMSSM fails to explain the diphoton and $b\bar{b}$ excesses and the $650~{\rm GeV}$ excess reported in~Ref.~\cite{CMS:2022tgk}, which corresponds to $\sigma_{\gamma \gamma b \bar{b}} = 0.35^{+0.17}_{-0.13}~{\rm fb}$~\cite{Ellwanger:2023zjc}, simultaneously.

In Ref.~\cite{CMS:2022goy}, the CMS collaboration presented the search for a new boson $\phi$ in $\tau \bar{\tau}$ final states, using the data samples collected in the full Run 2 phase of the LHC. It acquired $95\%$ confidence level bounds on the signal cross section, which was $12.2~{\rm pb}$ for $m_\phi = 95~{\rm GeV}$, corresponding to $\mu_{\tau \bar{\tau}}^{\rm exp} = 2.15$.  As discussed in Sec.~\ref{Section-excess}, the GNMSSM prediction of $\mu_{\tau \bar{\tau}}$ is around $0.11$ if one intends to explain the diphoton and $b\bar{b}$ excesses. It is much smaller than the bound. It is also significantly lower than the signal rate needed to explain the $\tau \bar{\tau}$ excess observed in~\cite{CMS:2022goy}, which is $\mu_{\tau \bar{\tau}} = 1.38^{+0.69}_{-0.55}$~\cite{Ellwanger:2023zjc}.

\section{Conclusion} \label{conclusion}
The CMS and ATLAS collaborations recently published their results searching for light Higgs bosons, using the complete Run 2 data of the LHC. Both reported an excess in the diphoton invariant mass distribution at $m_{\gamma \gamma} \simeq 95.4~{\rm GeV}$ with compatible signal strengths. These observations confirmed the excess previously reported by CMS, which was based on the analyses of the Run 1 data of the LHC and the first year of the Run 2 data. The combined result increased the local significance to $3.1\sigma$. Besides, the invariant mass of the diphoton signal coincided with that of the $b\bar{b}$ excess observed at the LEP. Although these excesses might originate from fluctuating much more extensive backgrounds, it is inspiring to speculate that they arise from the producing a CP-even Higgs boson with its mass around $95.4~{\rm GeV}$. If this thought proves true, it will be the first sign of new physics in the Higgs sector.

Given the remarkable theoretical advantages of the GNMSSM, we explained the excesses by the resonant productions of the singlet-dominated scalar, $h_s$, predicted by the theory. We proposed a new set of input parameters to acquire simple approximations of the signal strengths in a large $\tan \beta$ limit. With the help of these formulae, we learned the dependence of the excesses on the model parameters such as $\lambda$, $\mu_{tot}$, $m_A$, $m_B$, and $A_\lambda$. We also concluded that the central values of the signal strengths for the excesses corresponded to a moderately large SM Higgs field component in $h_s$ and a suppressed $h_s b \bar{b}$ coupling compared with the $h_s t \bar{t}$ coupling, i.e., $V_{h_s}^{\rm SM} \simeq 0.36$, $C_{h_s t \bar{t}} \simeq 0.36$, and $C_{h_s b \bar{b}} \simeq 0.25$. In particular, we showed that the small deviations of $C_{h_s g g}$ and $C_{h_s \gamma \gamma}$ from $C_{h_s t \bar{t}}$ could alleviate the suppression of the $h_s b \bar{b}$ couplings by significantly reducing the $H_{\rm NSM}$ component in $h_s$ and thus make the theory more accessible to explain the excesses. These observations guided us to find the parameter space responsible for the excesses.

We performed a sophisticated scan over the broad parameter space of the GNMSSM to investigate the impacts of various experimental restrictions, including those from the 125~{\rm GeV} Higgs data, the DM relic abundance and direct detection experiments, and the collider searches for SUSY and extra Higgs bosons, on the explanations. After analyzing the distributions of different parameters, we had the following conclusions:
\begin{itemize}
\item Present $125~{\rm GeV}$ Higgs data, and the collider searches for extra Higgs bosons were compatible with the existence of the light singlet-dominated Higgs boson responsible for the excesses. They influenced the Higgs physics only by setting lower bounds on the mass of charged Higgs bosons.
\item The DM physics could affect the explanations by determining the PLs and posterior PDFs of some parameters crucial for the excesses, such as $\lambda$ and $\mu_{tot}$. Specifically, the DM candidate might be $\tilde{B}$- or $\tilde{S}$-dominated $\tilde{\chi}_1^0$.  The $\tilde{B}$-dominated DM achieved the measured relic abundance mainly by co-annihilating with the wino-like electroweakinos. A small $\lambda$ and a sufficiently large $\mu_{tot}$ characterized this case in accounting for the excesses and simultaneously satisfying the restrictions from the DM direct detection experiments. Consequently, $V_{h_s}^{\rm NSM}$ was minor and the chargino's contribution to $C_{h_s \gamma \gamma}$ was never significant. By contrast, the $\tilde{S}$-dominated DM could obtain the correct abundance by co-annihilating with the higgsino-like or wino-like electroweakinos. This case allowed a larger $\lambda$ and a smaller $\mu_{tot}$ to provide a significant $V_{h_s}^{\rm NSM}$ and also enhance the supersymmetric contributions to $C_{h_s \gamma\gamma}$. As a result, the $\tilde{S}$-dominated DM case is slightly more suited to explain the excesses.
 \item The GNMSSM primarily relied on the Higgs mixings instead of significant supersymmetric contributions to $C_{h_s g g}$ and $C_{h_s \gamma \gamma}$ to explain the excesses. Given that this mechanism could be realized even for massive sparticles, the explanation possessed broad parameter space consistent with the LHC searches for SUSY by either Strategy I or II. We verified this point by simulating some samples expected to have distinguished signals at the LHC.
 \item Given the parameter space of the GNMSSM in Table~\ref{tab:1}, the signal strengths $\mu_{\gamma \gamma}$ and $\mu_{b\bar{b}}$ could reach 0.36 and 0.25 without conflicting the experimental restrictions. Their ratio varied from 1.25 to 2.0 for the $\tilde{B}$-dominated DM case and 1.25 to 2.3 for the $\tilde{S}$-dominated DM case. Consequently, the GNMSSM could simultaneously explain the diphoton and $b\bar{b}$ excesses at the $1\sigma$ level. In particular, the $\tilde{S}$-dominated DM case could predict the central values of the signal strengths for the excesses.
\end{itemize}

We add that our explanation predicts the cross sections of the $\gamma\gamma b\bar{b}$ signal from process $pp \to H \to h_s h$ at the LHC and the $\tau \bar{\tau}$ signal from the production $ p p \to h_s \to \tau \bar{\tau}$ in simple forms that are consistent with corresponding experimental bounds. We look forward to seeing that the run 3 results from ATLAS and CMS and future runs of the high luminosity LHC could illuminate whether the excesses persist and arise from a BSM particle. We expect that future linear colliders could provide definite conclusions on these excesses.

\section*{Acknowledgements}
We thank Dr. Junquan Tao for helpful discussions about the details of the diphoton excess observed by the CMS collaboration. This work is supported by the National Natural Science Foundation of China (NNSFC) under grant No. 12075076.

%\bibliographystyle{unsrt}
%\bibliographystyle{JHEP}
%\bibliography{SD}

\bibliographystyle{CitationStyle}
\bibliography{LianRef}

\end{document}